\documentclass[]{tPHM2e}
\usepackage{graphicx}
\usepackage{epsfig}

\usepackage{color}

\newcommand{\urs}{URu$_2$Si$_2$}

\begin{document}
\doi{10.1080/14786435.20xx.xxxxxx}
\issn{1478-6443}
\issnp{1478-6435}
\jvol{00} \jnum{00} \jyear{2014} \jmonth{May}


\articletype{Article in Special Issue}


\title{{\itshape  Hidden Order Behaviour in URu$_2$Si$_2$} \break(A Critical Review of the {Status of} Hidden Order in 2014)}  

\author{J. A. Mydosh$^{\rm a}$$^{\ast}$\thanks{$^\ast$Corresponding author. Email: mydosh@physics.leidenuniv.nl
\vspace{6pt}} and P. M. Oppeneer$^{\rm b}$\\\vspace{6pt}  $^{\rm a}${\textit{Kamerlingh Onnes Laboratory, Leiden University, PO Box 9504, NL-2300 RA Leiden, The Netherlands}} and 
$^{\rm b}${\textit{Department of Physics and Astronomy, Uppsala University, Box 516, S-75120 Uppsala, Sweden,}}\\\vspace{6pt}\received{April 2014} 
}

\maketitle

\begin{abstract}
Throughout the past three decades the hidden order (HO) problem in URu$_2$Si$_2$ has remained a ``hot topic" in the physics of strongly correlated electron systems with well over 600 publications related to this subject. Presently in 2014 there has been significant progress in combining various experimental results embedded within electronic structure calculations using density functional theory (DFT) to give a consistent description of the itinerant behaviour of the HO transition and its low temperature state. Here we review six different experiments: ARPES, quantum oscillations, neutron scattering, RXD, optical spectroscopy and STM/STS. We then establish the consistencies among these experiments when viewed through the Fermi-surface nesting, folding and gapping framework as predicted by DFT. We also discuss a group of other experiments (torque, cyclotron resonance, NMR and XRD) that are more controversial and are presently in a ``transition" state regarding their interpretation as rotational symmetry breaking and dotriacontapole formation. There are also a series of recent ``exotic" experiments (Raman scattering, polar Kerr effect and ultrasonics) that require verification, yet they offer new insights into the HO symmetry breaking and order parameter. We conclude with some constraining comments on the microscopic models that rely on localised $5f$-U states and strong Ising anisotropy {for explaining} the HO transition,  
{and with an examination of different models in the light of recent experiments.}
\end{abstract}
 
 \date{today}

\bigskip

\centerline{\bfseries Index}
\hbox to \textwidth{\hsize\textwidth\vbox{\hsize18pc}}
\hspace*{-12pt}


{1.}  Introduction

{2.}  ARPES and quantum oscillations {in comparison} with DFT

{3.}  Neutron scattering, XRS, and DFT and crystal symmetry  

{4.}  STM/STS, optical conductivity, and DFT 

{5.}  Controversial experiments and theory  

{6.}  New ``exotic" experiments and relation to theory 

{7.}  Localized versus itinerant descriptions 

{8.} {A confrontation of recent models}

{9.}   Conclusions

{10.}Ê{~Acknowledgements}

\section{Introduction}
In 1984 a remarkable discovery was made with the observation of superconductivity in an ``antiferromagnetic" uranium-based heavy-fermion compound
\cite{pals85,mapl86,schl86}.  {Now,} 30 years later {it has become realised} that the superconductivity at 1.5~K was of the unconventional, {time-reversal symmetry violating}, $d$-wave type \cite{kasa07,liii13}. However,  the putative antiferromagnetism was not of magnetic nature but of a new and unknown type of non-magnetic, non-structural order,  thus the term \textit{hidden order} was coined. Initially, enormous efforts were expended in studying the superconductivity  and mysterious HO via the standard bulk type of measurements \cite{myds11}. It was only when neutron scattering was brought to bear on the issue that  the HO became topical and transmuted by pressure into a true local-moment antiferromagnet  (see \cite{myds11} and references therein).  The large entropy change of the HO phase transition at T$_0 =17.5$~K was not a {conventional dipolar} spin order or a crystal structure modification. 
{So now research has progressed}
into 2014 {with hope} to finally glimpse the novel order parameter, its symmetry, and the corresponding excitations via a viable microscopic theory. 

Table 1 lists the contemporary, in progress, experiments of the most modern ``cutting edge" methods  {often} for the first time being applied to {the} heavy-fermion URu$_2$Si$_2$ [7--44]. This material serves as a test bed for novel heavy-fermion exploration with the use of exotic and expensive techniques. Here in Table 1 we have neglected the many standard ``bulk" thermodynamic, transport and magnetic experiments performed since 1985 \cite{myds11}. In what follows we will attempt to interrelate a number of these experiments and illustrate them within the DFT band structure and Fermi surface (FS) behaviour. Other experiments lie outside this regime and we will consider them as controversial. In addition very recent ``exotic" experiments are listed in the figure and summarised in Sec.~6. Many of such measurements are provocative, shedding new light on the symmetry breaking and order parameter of the HO state, yet they await full confirmation and interpretation.

We compile in Table 2 a section of the large number of theoretical efforts [45--61] to interpret the HO phase transition through analytic models and calculations. Note how these have spanned almost 20 years of work. Yet there is no real consensus  as to the correct one. In addition to such theories, Table 3 presents the many proposals [62--73] for a multipolar HO transition, beginning with the modified dipole moment antiferromagnetism and extending to rank-5 dotriacontapole. No one as yet has proposed a monopole. For a further updating into 2014 of the theoretical efforts and references, see Refs.\ \cite{chan13} and \cite{khal14}. 

Full explanation, critique and possible validation of theories listed in Tables 2 \& 3  are beyond the scope of this review. We will only briefly discuss a relevant theory according to its relation to the given experiment. Finally, we treat the debate over the localised description that relies on the experimental realisation of  a strong Ising-like anisotropy in URu$_2$Si$_2$ that even occurs in dilute Th$_{1-x}$U$_x$Ru$_2$Si$_2$.  {The huge Ising anisotropy of quasiparticles in {\urs} has been emphasised as a hallmark of localised $5f$ character \cite{alta12,chan13}. }{Notwithstanding,} here we will illustrate that an itinerant DFT description can also explain the Ising anisotropy. {Continuing along this line, we examine several recent itinerant models.}

\section{ARPES \& quantum oscillations {in comparison} with DFT} 

\begin{figure}
\begin{center}
\includegraphics[width=5in]{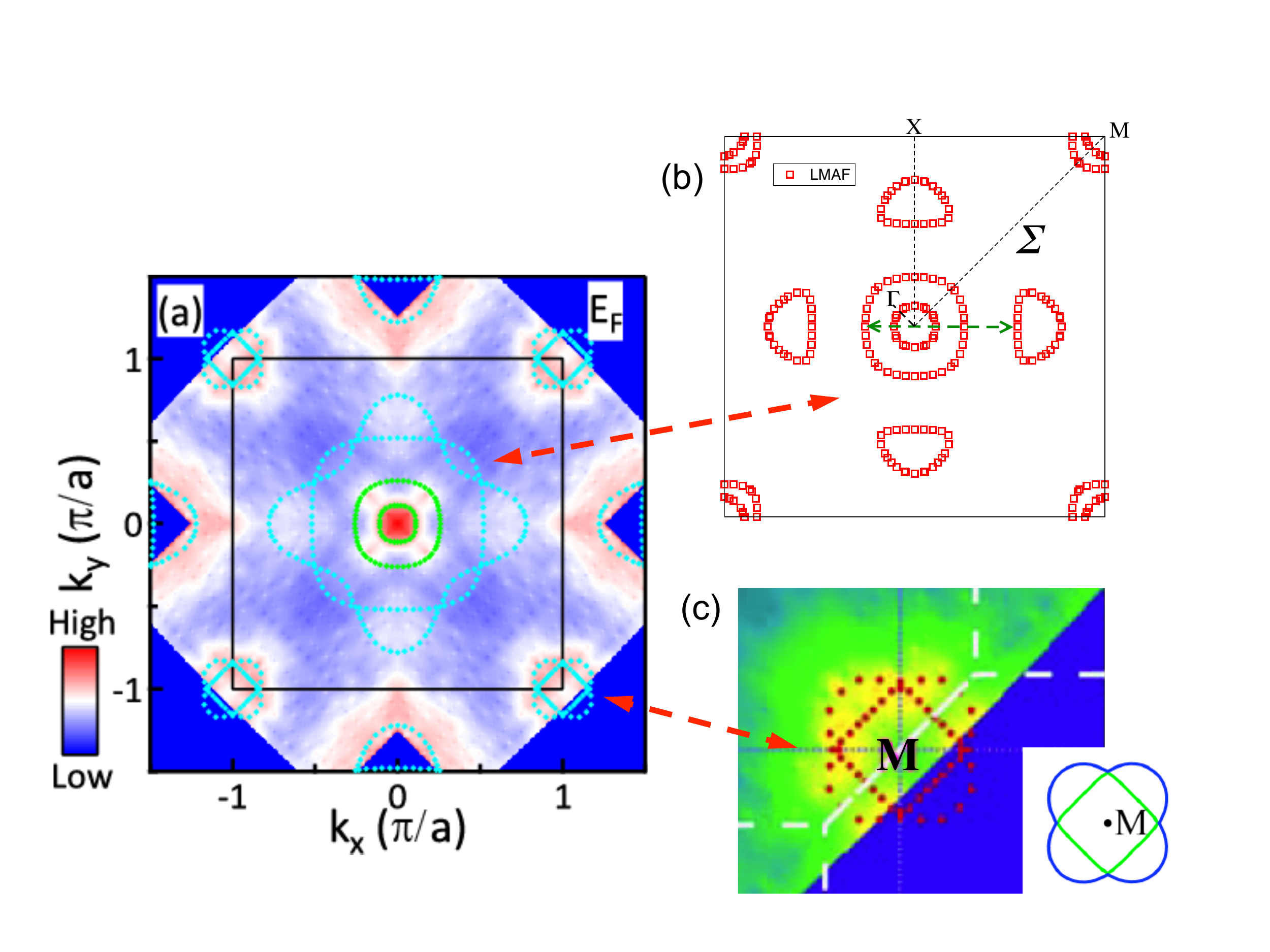}
\caption{(a) ARPES spectral weight measured in the  $k_z =0$ ($\Gamma$) plane at the Fermi energy ($E_{\rm F}$) for T=12~K, after Meng et al.\ \cite{meng13}.  The superimposed DFT calculated FS cross-sections (light blue lines) \cite{elga09}  for the nonmagnetic phase are also shown. (b) DFT computed FS cross-sections of the LMAF phase in the simple tetragonal Brillouin zone. Note the lack of intensity in the [110] ($\Sigma$) directions indicating the opening of an energy gap. (c) Enlarged view of the ARPES intensity around the M point (yellow), with superimposed DFT cross-section (red dotted lines). The structure reveals two overlaying pockets, evidencing thus downfolding over $\boldsymbol{Q}_0 = [0\, 0\, 1]$ in the HO. } 
\label{fig1}
\end{center}
\end{figure}

There has recently been a surge of {angular-resolved photoemission spectroscopy} (ARPES) experimentation on URu$_2$Si$_2$. The first {ARPES measurements which surveyed the} temperature dependence {upon moving} through the HO transition were {made by} Santander-Syro et al. \cite{sant09}. {They} claimed {to observe} the  {traversal} of a band of heavy quasiparticles through the Fermi level {at T$_0$}. A series of extended ARPES measurements followed where now the entire Brillouin zone has been spectroscopically covered as the temperature is varied both above and below T$_0$. The HO state \cite{meng13} is clearly characterised by a restructuring of the FS at certain regions of the 3D Brillouin zone. Thus, the {emerging picture is that the} HO transition is related to a partial gapping of the FS. Figure \ref{fig1} shows a representative picture of the recent Fermi surface ARPES spectroscopy. For the sake of comparison the DFT band structures are superimposed. Note the good relationship between experiment and theory. For additional details, see the complete ARPES treatment in this special issue by T. Durakiewicz et al. 

\begin{figure}
\begin{center}
\includegraphics[width=3in]{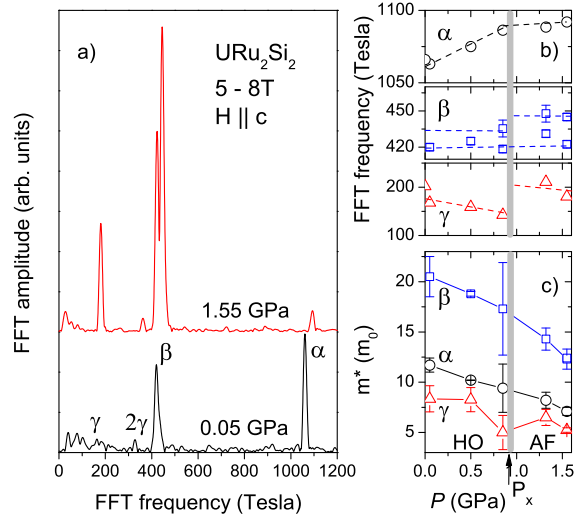}
\caption{Shubnikov-de Haas measurements of quantum oscillations in URu$_2$Si$_2$  at 35 mK. (a)  Fast Fourier transforms (FFT) of the S-dH oscillations showing the FFT amplitude  vs.\ FS extremal area frequencies at two different pressures and magnetic fields of 5 to 8~T. (b) Pressure dependence of the FFT frequencies, and (c) of the effective masses $m^{\star}$. Note the limited modifications in both quantities upon crossing the critical (HO to LMAF) pressure, P$_x$. After Hassinger et al.\ \cite{hass10}.}
\label{fig2}
\end{center}
\end{figure}

Initial de Haas-van Alphen (dH-vA) and Shubnikov-de Haas (S-dH)  {data} have been available onwards from {pioneering work in} 1999 \cite{ohku99} onwards with efforts continuing into the present \cite{alta12}. The quantum oscillation (QO) {experiments} are limited to very low temperatures at significant ranges of applied magnetic field. With great effort, many FS orbits have been observed and favourably compared not only with the ARPES FS but also with the DFT  FS  {computed}  for the LMAF state. By way of improved experimentation \cite{hass10} a new variable, viz., hydrostatic pressure, is now available to track the HO into the large-moment antiferromagnetic (LMAF) phase. The key result here is that, {surprisingly,} the Fermi surface is minimally modified between the two phases HO and LMAF as shown in Fig.\ \ref{fig2} (from \cite{hass10}). Most importantly, this non-mutated FS then justifies the use of antiferromagnetic DFT Fermi surface calculations transposed into the non-magnetic HO phase. 

\begin{figure}
\begin{center}
\includegraphics[width=1.8in]{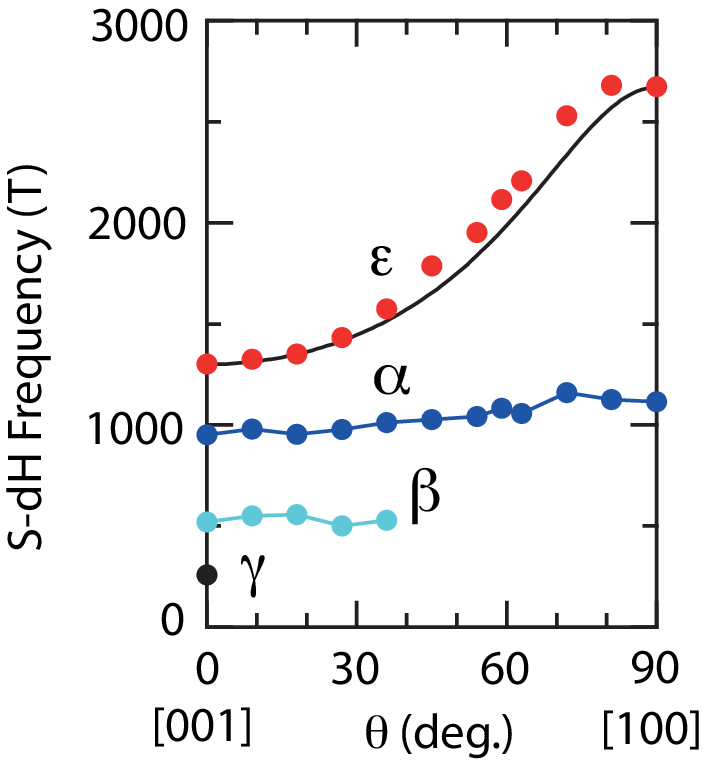}
\includegraphics[width=1.8in,height=2.15in]{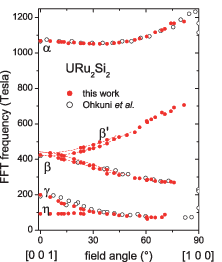}\\
\includegraphics[width=1.45in]{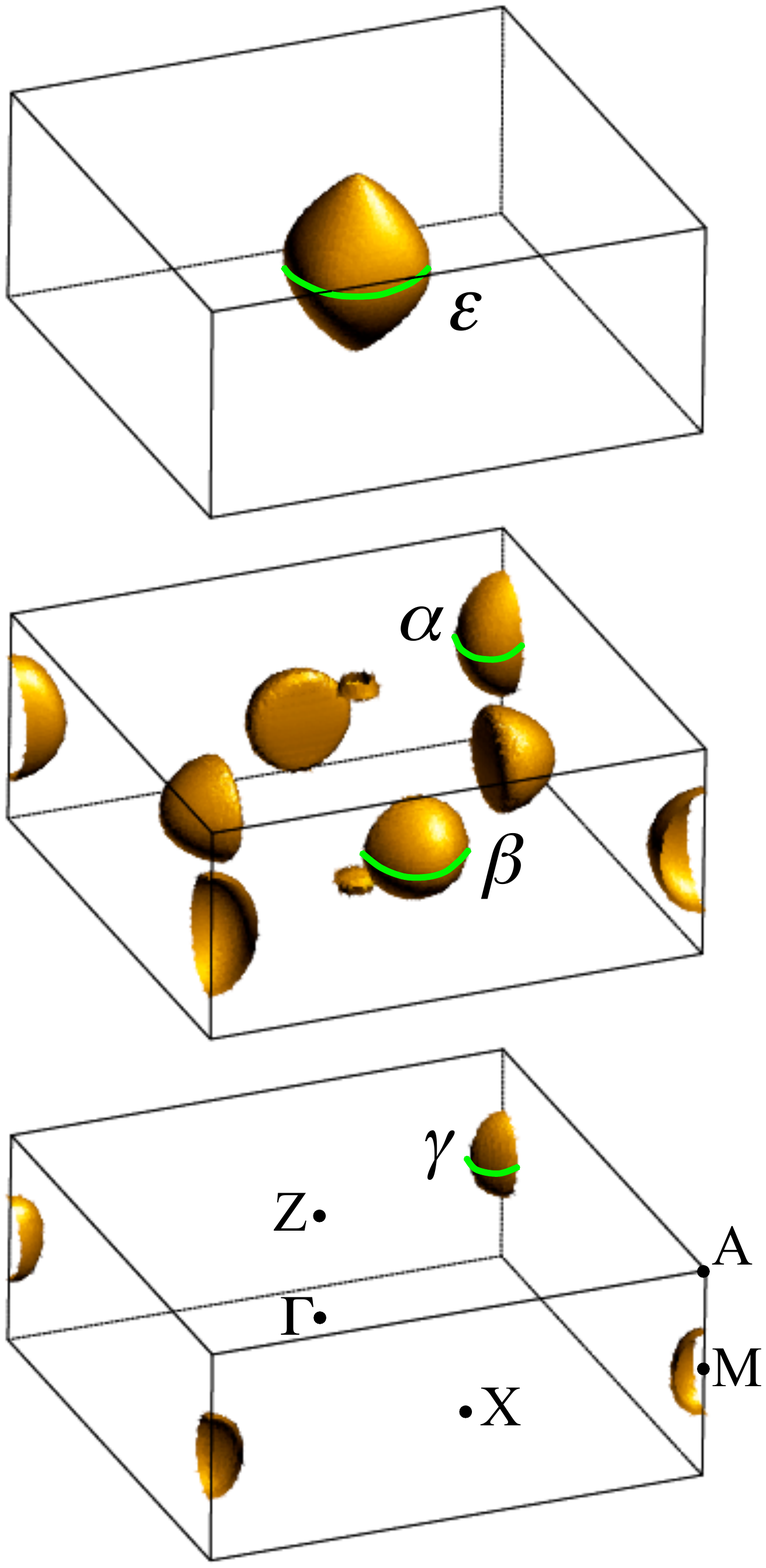}
\includegraphics[width=2.5in]{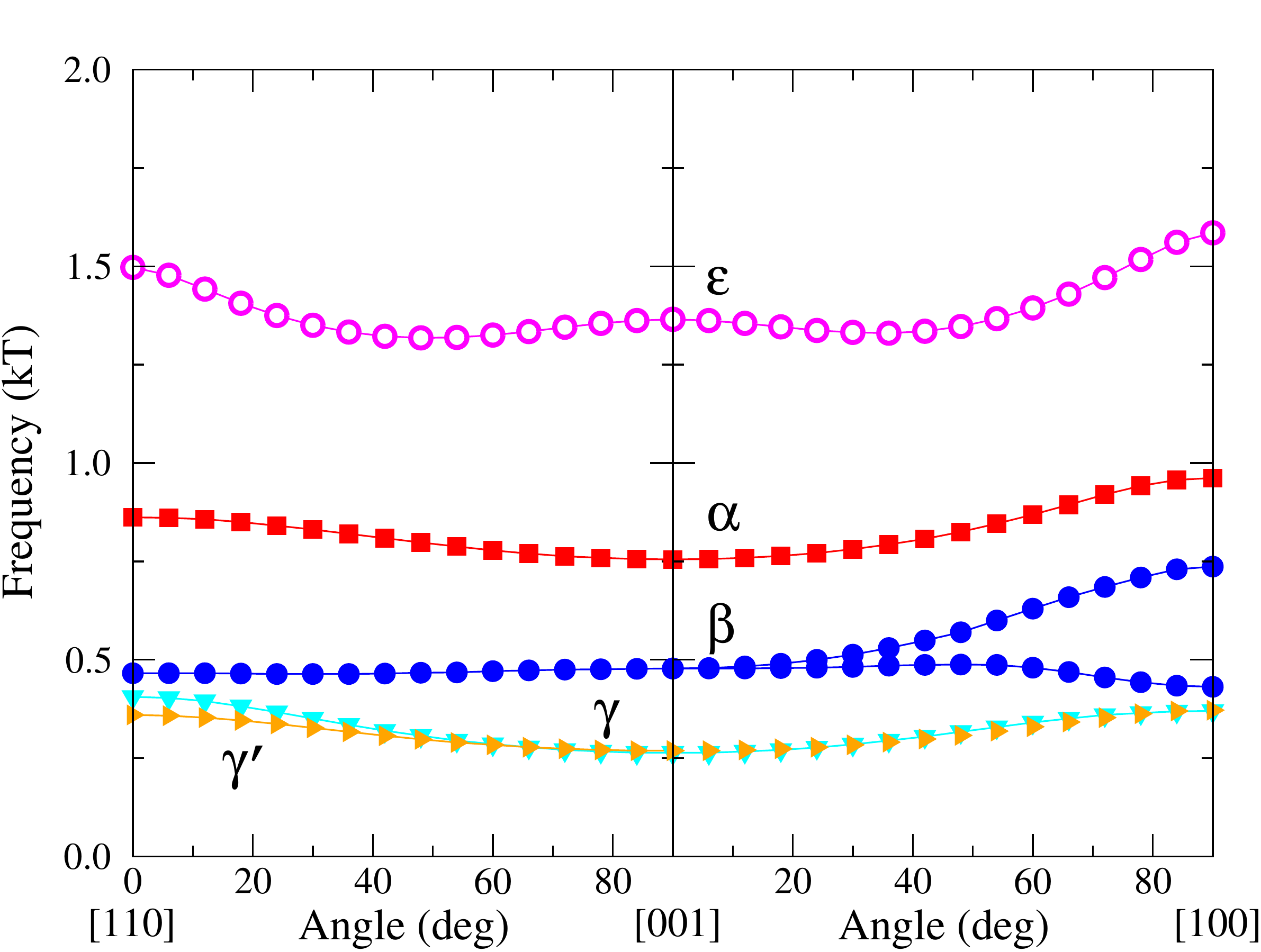}
\caption{Top panels: Angular dependence of various FS branches, obtained from S-dH quantum oscillation measurements (left panel from Shishido et al.\ \cite{shis09}, right panel from Hassinger et al.\ \cite{hass10}). Bottom panels: The DFT calculated  FS (left) and the DFT calculation of the angular dependence of the extremal branches \cite{oppe11}. There is an overall good agreement between QO measurements and DFT calculations.}
\label{fig3}
\end{center}
\end{figure}

As a further check of the efficacy of the DFT {band structure model} we draw a comparison between the angular dependence of the orbits determined by S-dH and those calculated by DFT. Figure \ref{fig3} shows that the experimental--theoretical agreement for a number of orbits is excellent \cite{oppe10}. {We note, however, that there are different opinions on what is to be identified as the $\varepsilon$ and $\alpha$ branches \cite{tone12}.} {Nonetheless,} {it appears that} we are now reaching an effective correspondence among the detailed ARPES and QO 
results and the DFT calculations. This strongly indicates that the HO phenomenon is {best seen as a} gapping of {dispersive} U-${5f}$ bands near the FS at certain specific regions of the Brillouin zone. {To reach this result ARPES} measurements throughout all of ${k}$-space {are required}, a previous difficulty that has now been eliminated.


\section{Neutron scattering, XRS; DFT and crystal symmetry} 

\begin{figure}
\begin{center}
\includegraphics[width=3in]{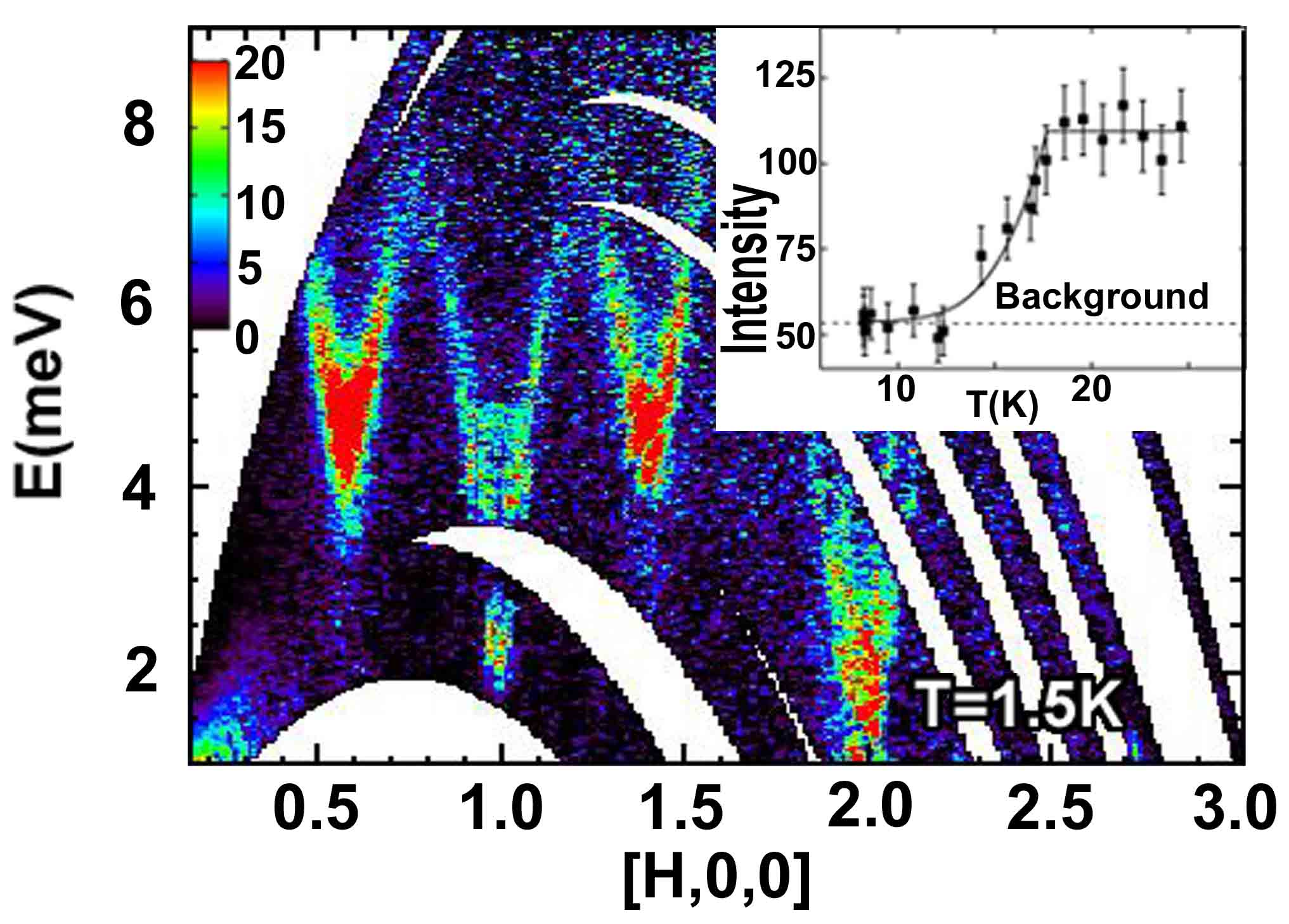}
\caption{Intensity plot of inelastic neutron scattering at T$=1.5$ K. Bright colours give the reciprocal space cone-like dispersions of two sharp resonant modes at $\boldsymbol{Q}_0=[1,0,0]$ and at $\boldsymbol{Q}_0=[1\pm0.4,0,0]$.  After Wiebe et al.\ \cite{wieb07}.}
\label{fig4}
\end{center}
\end{figure}

\begin{figure}
\begin{center}
\includegraphics[width=3in]{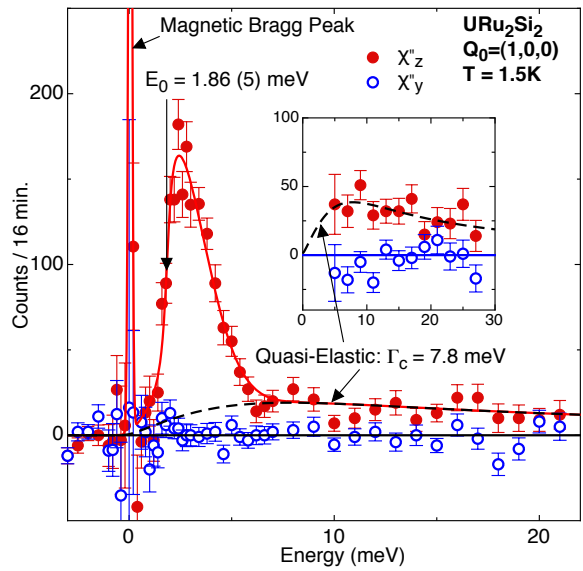}
\caption{Inelastic neutron scattering measurement of the commensurate longitudinal magnetic resonance at $\boldsymbol{Q}_0$ at 1.5~K. Apart from the sharp resonance there is a quasi-elastic continuum persisting to higher energies. After Bourdarot et al. \cite{bour10}.}
\label{fig5}
\end{center}
\end{figure}

\begin{figure}
\begin{center}
\includegraphics[width=4in]{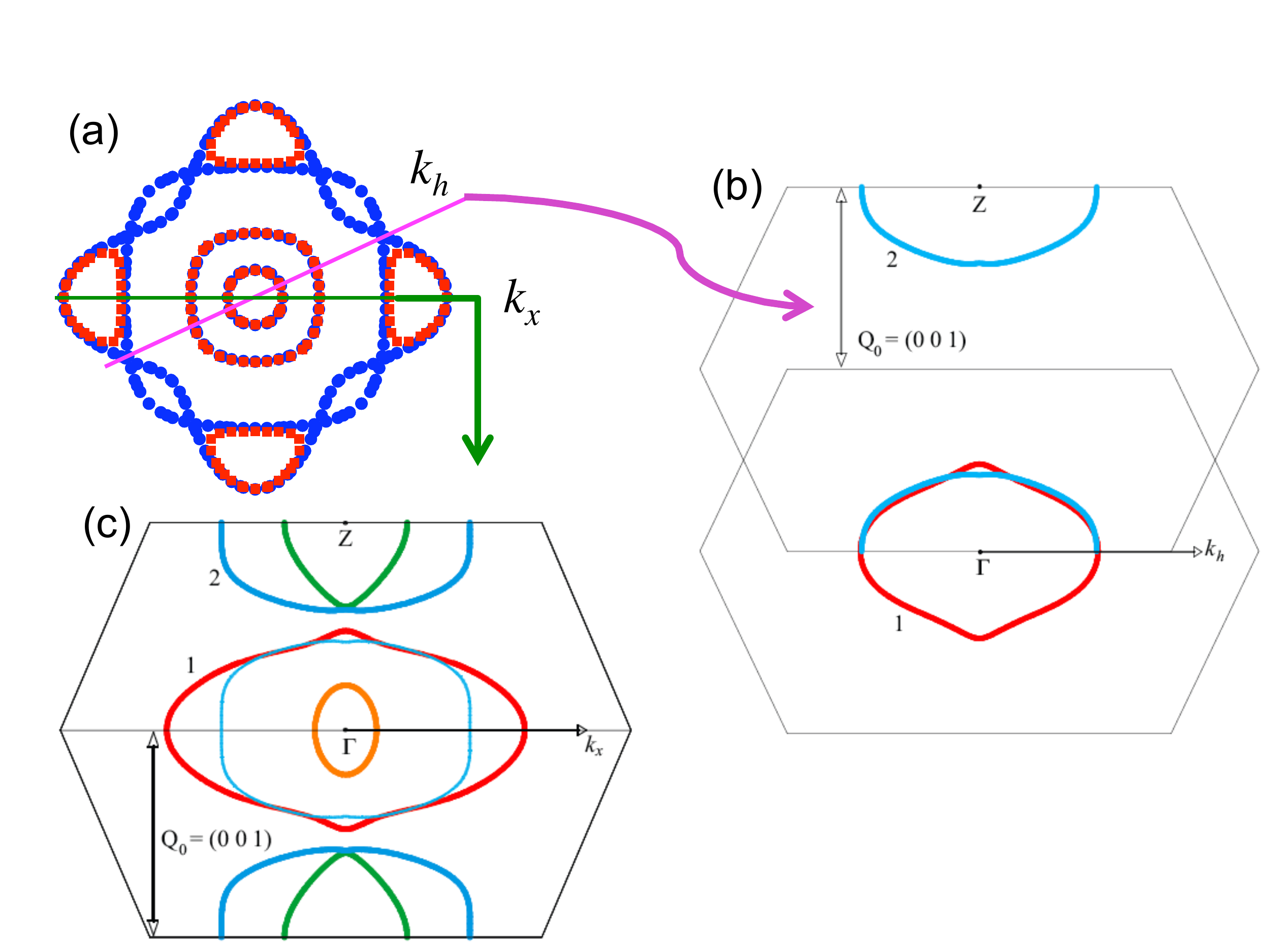}
\caption{Fermi surface nesting of two bands labeled 1 and 2 in the bct Brillouin zone. (a) FS cross-section in the $k_z =0$ plane with ``hot spot" line indicated by $k_h$ and high-symmetry line indicated by $k_x$. (b) A Brillouin zone  cross-section of $k_h - k_z$ shows that nearly perfect nesting over $\boldsymbol{Q}_0$ of the two FS sheets is present at the low-symmetry hot spot line. (c) A $k_x - k_z$ cross-section shows that a poorer nesting of the two sheets occurs along the high-symmetry $k_x$ direction. After Oppeneer et al.\ \cite{oppe11}.}
\label{fig6}
\end{center}
\end{figure}

It was the early neutron diffraction \cite{broh87,broh91} that initiated particular interest in {\urs} and its unusual HO. Already here we learned that (i) The magnetic {dipole} scattering (most probably extrinsic) was {much} too small to explain the large entropy {release at the} HO {phase} transition, (ii) There were two unusually sharp {inelastic neutron scattering} (INS) modes in the HO  state emerging out of {longitudinal} spin fluctuations {appearing} at higher temperatures and (iii) No non-singlet crystal electric field (CEF) levels could be detected. Figures \ref{fig4} \& \ref{fig5} exhibit a modern version of the neutron dispersion illustrating these two INS modes \cite{wieb07,bour10}. The commensurate $\boldsymbol{Q}_0=[0,0,1]$ mode is related to the strong FS nesting from ${\Gamma}$ to Z, {appearing in the body-centred tetragonal normal phase above T$_0$.} The nesting and folding (translating Z downwards {to $\Gamma$} by [0,0,1]) has been predicted by DFT as is shown in Fig.\ \ref{fig6} with FS projected in the basal plane \cite{oppe11}. It is this strong nesting {of the two FS sheets which is believed to} cause the HO gapping. Note that this [0,0,1] mode becomes an elastic Bragg peak when pressure is used to tune the HO to the LMAF  {phase}. The incommensurate $\boldsymbol{Q}_{1}$=[1.4,0,0] mode is {believed to be} due to the weaker nesting {in the $\Gamma$ plane \cite{elga09}}  that is induced by the strong spin-orbit coupling as found by DFT that generates a spin-orbit density wave \cite{tdas14}. $\boldsymbol{Q}_1$, which remains inelastic in the LMAF phase, is the basis of the spin fluctuations above T$_0$ that because of its gap in HO leads to the removal of the large entropy \cite{wieb07}. Therefore, with the DFT these two INS modes can be given a physical significance as consequences of the HO state.    
 
After these initiating studies numerous   {expanding} neutron experiments followed. One of which was a {recent} study of the magnetisation distribution induced by a $c$-axis magnetic field (10~T) using the  elastic scattering of polarised neutrons \cite{ress12}. Various analyses were employed to interpret the subtle shape changes (T$>$T$_0$ to T$<$T$_0$) of the magnetisation {density} surrounding the U-atoms. The analysis within  {a} localised ${5f^2}$ ionic model led to {the supposed} fingerprints of a magnetic dotriacontapole (rank-5) {as origin of the HO}. The totally normal form factor of the scattering intensity was {however} not considered; {earlier measurements of the neutron form factor} already indicated the {absence of a magnetic multipole}  \cite{kuwa06}. {Supported by} DFT calculations of Ikeda et al.\ \cite{iked12} {which predicted a magnetic dotriacontapole,} the {announcement} of ``unveiling'' the HO as the freezing of rank-5 multipoles was made. Very recently, {an} examination \cite{khal14} of the magnetic space-groups for URu$_2$Si$_2$ was employed to {study the magnetic form factor and analyse}  the exact properties of the HO order parameter using the available magnetic neutron scattering amplitude. This symmetry analysis, {however,} discounted the dotriacontapole and all parity-even, time-odd multipoles. Because the rank-2 {non-magnetic} quadrupole had been previously eliminated {through x-ray resonant scattering (XRS)} \cite{walk11},  only the parity-odd, time-even (non-magnetic, rank-4) hexadecapole remains {as a possible source for the HO.} Since neutron scattering is ineffective here, resonant x-ray Bragg diffraction is required. Future efforts with these  {demanding} synchrotron multipole techniques are planned \cite{walk14}.

\section{STM/STS, optical conductivity and DFT} 

\begin{figure}
\begin{center}
\includegraphics[width=3in]{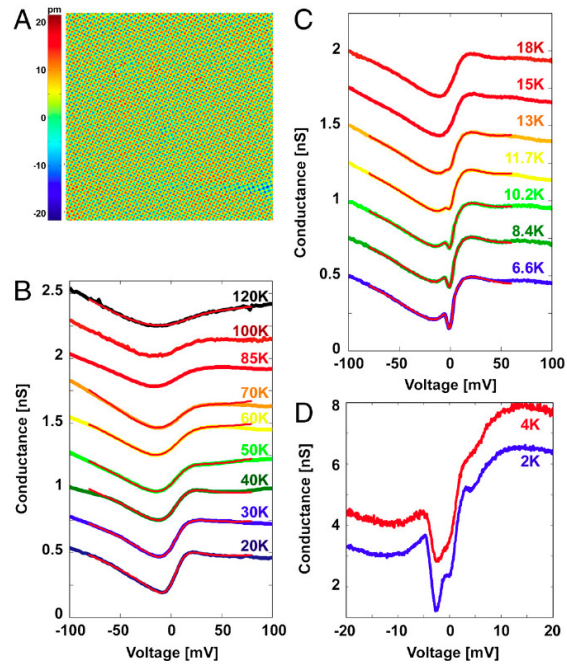}
\caption{(A) Topographic STM image showing atomically ordered Si-terminated surface. (B) Tunneling spectroscopy conductance vs.\ voltage measured for several temperatures above T$_0$, and (C) below T$_0$.
(D) The averaged conductance at low temperatures showing that the electronic density of states exhibits additional features in the HO gap. After Aynajian et al.\ \cite{ayna10}.}
\label{fig7}
\end{center}
\end{figure}

In the last few years scanning tunnelling microscopy and spectroscopy (STM/STS) {measurements} have become available {assisted by developments of} the experimental procedure of \textit{in-situ} cleaving a ``perfect"  surface of URu$_2$Si$_2$ as an art of the science. Two such studies  appeared in 2010 \cite{ayna10,schm10} on URu$_2$Si$_2$ as the first wide temperature scans of a heavy-fermion material. These dramatic  results inaugurated the investigations of  strongly  correlated intermetallic compounds via STM/STS and at present many new efforts have followed \cite{ayna14}. In Fig.\ \ref{fig7} we collect the initial measurements of Aynajian et al. \cite{ayna10}, the first at high temperature, with the slow opening of the hybridisation gap describable as a Fano resonance. The asymmetric Fano-like lineshape is caused by the two channel tunnelling into the U-layer due to the surface termination of the cleave. The two channels {are} a result of quantum interference between a discrete or local state (t$_f$) and a continuum of conduction electron (t$_c$) states. For URu$_2$Si$_2$ in Fig.\ \ref{fig7} there is a mixture of the two: t$_f$ results from  heavy-fermion bands that are formed well above T$_0$. The energy scale for this hybridisation crossover is ca.\ 20 meV. A similar behaviour is found in {the heavy-fermion} CeCoIn$_5$ \cite{ayna14} which represents the generic formation of the {coherent} HF state through hybridisation {of incoherent local $f$ states and conduction Bloch electron states}. At lower temperatures (see Fig.\ \ref{fig7}) STS tracks the sudden in temperature opening of the HO gap. $\Delta_{HO}$(T) follows a second-order, phase-transition behaviour that can be approximated by a BCS-gap temperature dependence, $\Delta_{HO} (0) = 5$ meV. Note {that} there are modes or states within the HO gap that appearer at yet lower temperatures. Accordingly  STM/STS clearly indicates two distinctly different energy gaps: hybridisation and HO. Unfortunately, the third, superconducting gap below T$_c \approx 1.5$~K has not yet been studied.

\begin{figure}
\begin{center}
\includegraphics[width=2.3in]{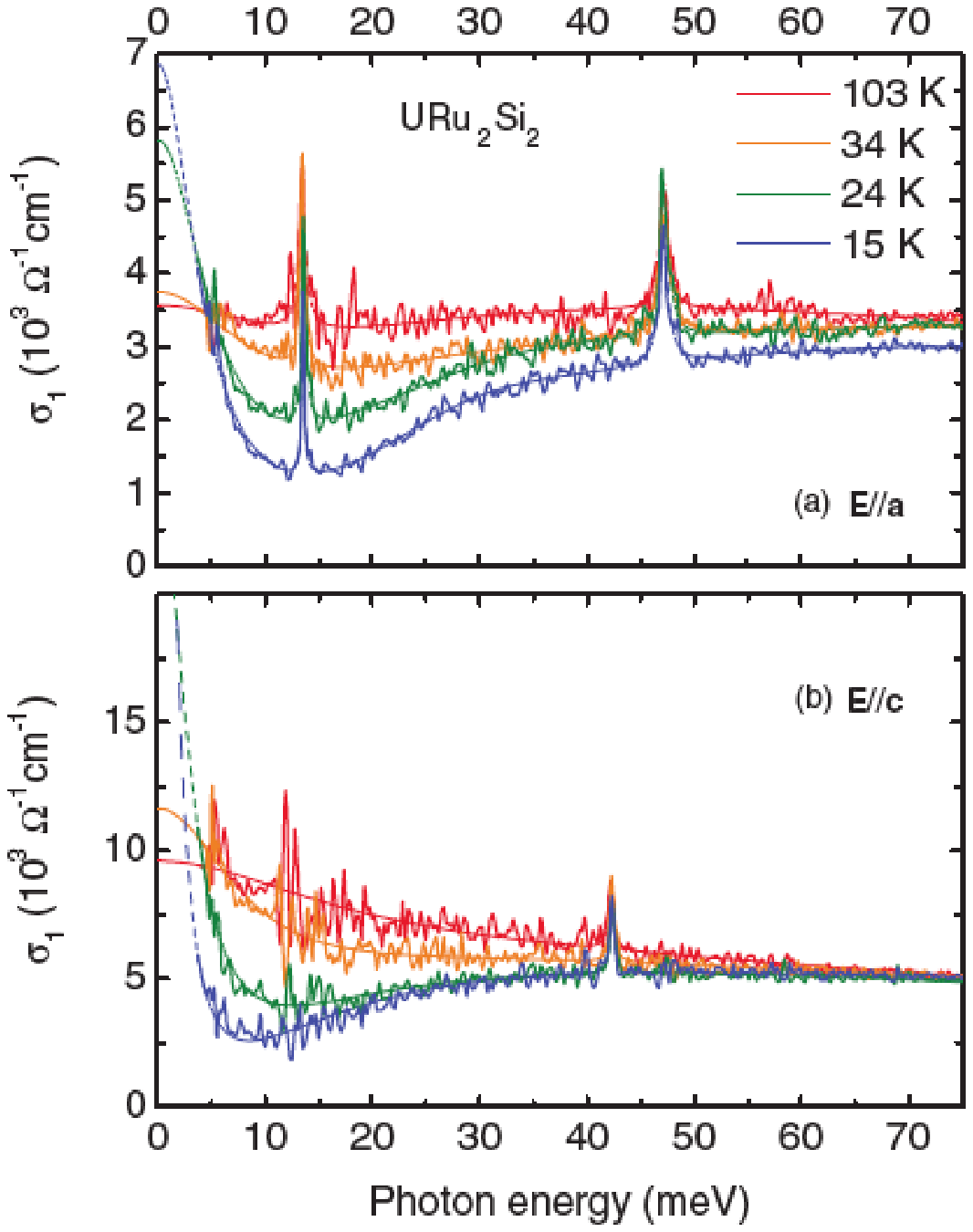}
\includegraphics[width=2.7in]{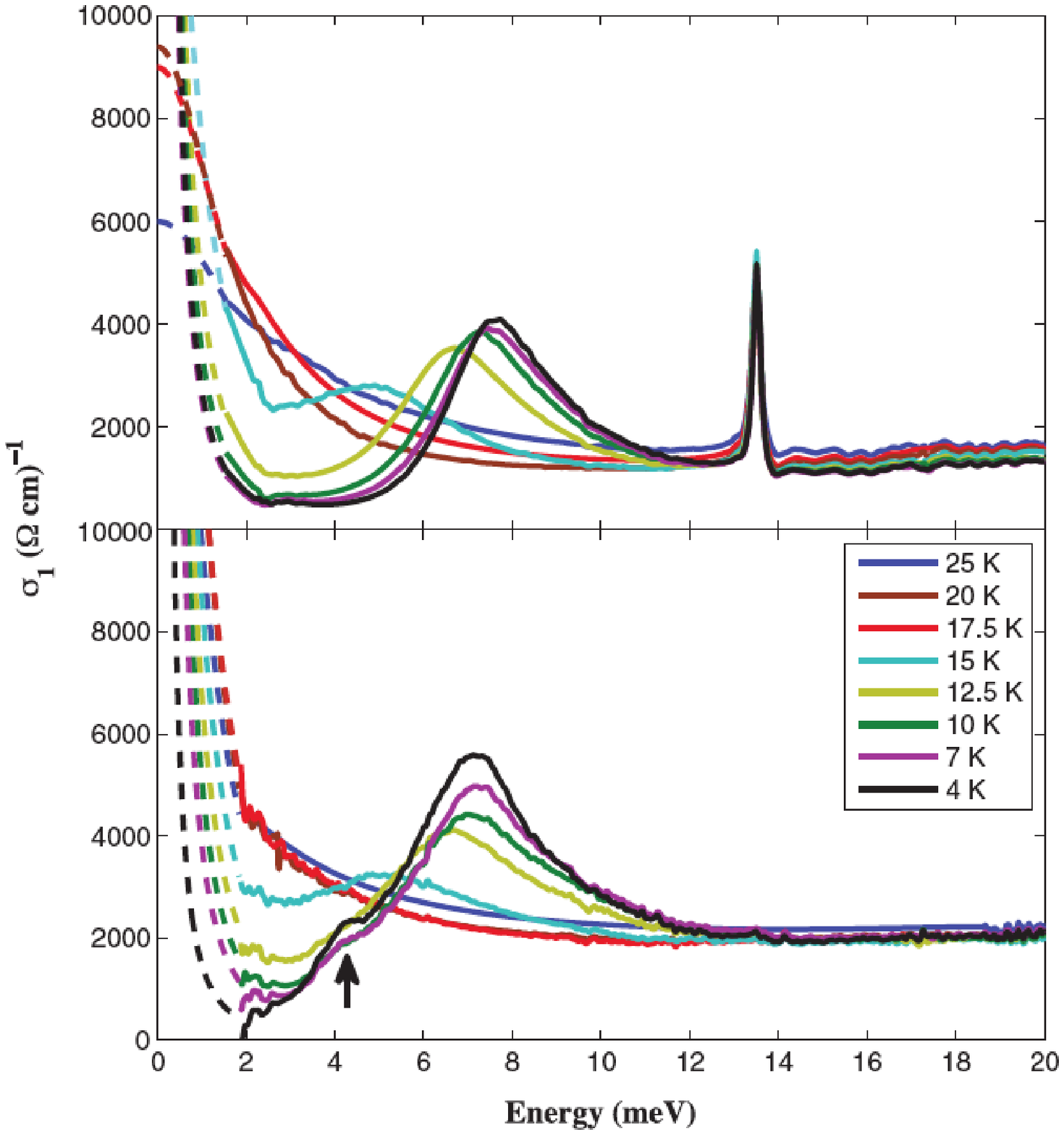}
\caption{Optical conductivity spectrum measured for several temperatures and with light polarisation $E\,||\,a$ and $E\,||\,c$. Left panels: high-temperature results from Levallois et al. \cite{leva11}. Right panels:
low temperature results from Hall et al.\ \cite{hall12}. Top panel is for $E\,||\,a$ and bottom panel for $E\,||\,c$.}
\label{fig8}
\end{center}
\end{figure}

We remark that STM/STS is a surface technique that must be justified as detecting bulk effects. In contrast,  optical spectroscopy directly  {provides} a bulk determination. Investigations of optical spectroscopy (OS) were begun in 1988 by Bonn et al. \cite{bonn88}. Although their data are sparse they were able to clearly show the opening of the HO gap below 17~K of $\sim 6$ meV and indications for the hybridisation gap above T$_0$ of $ \sim 15 - 20$ meV. Recent work has confirmed these effects with systematic low frequency and temperature data. Figure \ref{fig8} displays the anisotropic slow  opening of the hybridisation gap for optical conductivity measurements parallel to the ${a}$ and ${c}$ axes \cite{leva11}. Note there are clear observations of hybridisation effects at 34~K and anisotropy in the gap structure as a function of energy. Additional experiments that confirm the existence of a hybridisation gap above T$_0$ as distinct from a HO gap are given by the OS of Guo et al. \cite{guoo12}. 
{The hybridisation gap is sometimes related to a different gap, a pseudogap \cite{shir12}, which has been proposed to occur below about 25~K  \cite{hara11}. Such pseudogap was suggested by recent PCS measurements \cite{park12} that detected the onset of gap below 27~K, but could hardly observe any notable change at T$_0$. Recent PCS studies using a soft contact however detected only a change in the differential conductance with an onset at T$_0$ \cite{luuu12}.}

Figure \ref{fig8} shows the real part of the optical conductivity within the temperature region 4 to 25~K with measured data down to 2 meV (16 cm$^{-1}$) and with the light polarised along ${a}$ and ${c}$ axes \cite{hall12}. As is seen the evolution of the HO gap is preceded (as $\omega $ is decreased) by a density-wave-like maximum. The extrapolation of the Drude peak (dashed lines) to ${\omega} \rightarrow  0$ is estimated by the dc conductivity end point. The interesting result here is the observation of a multiple-gap structure in the HO phase, the values of which correspond roughly to those found from other techniques, e.g., STS and neutron scattering, 3 to 5 meV. The origin of the multi-gap HO formation is unclear at present but it is certainly a bulk effect. 
            
Nevertheless, both of these distinctly different (surface vs.\ bulk) techniques lead to the same global conclusion. Namely, the HF state is formed by electron-electron interactions leading to hybridisation {and incorporation} of the magnetic spins into a heavy Fermi liquid with loss of local magnetic moments, {at a temperature considerably above T$_0$}. 
Or complementary, there is the Kondo-lattice compensation which blots out the  {local magnetic moment.} Such behaviour appears {generic} in the many HF materials and acts as a precursor to the generation of a unique ground state. For 
URu$_2$Si$_2$ this state is the HO phase formed mysteriously out of the heavy-Fermi liquid. Therefore, STS and OS nicely agree with each other and are {both} related to the DFT FS and gapping predictions. 
   

\section{Controversial experiments and theory} 

\begin{figure}
\begin{center}
\includegraphics[width=4in]{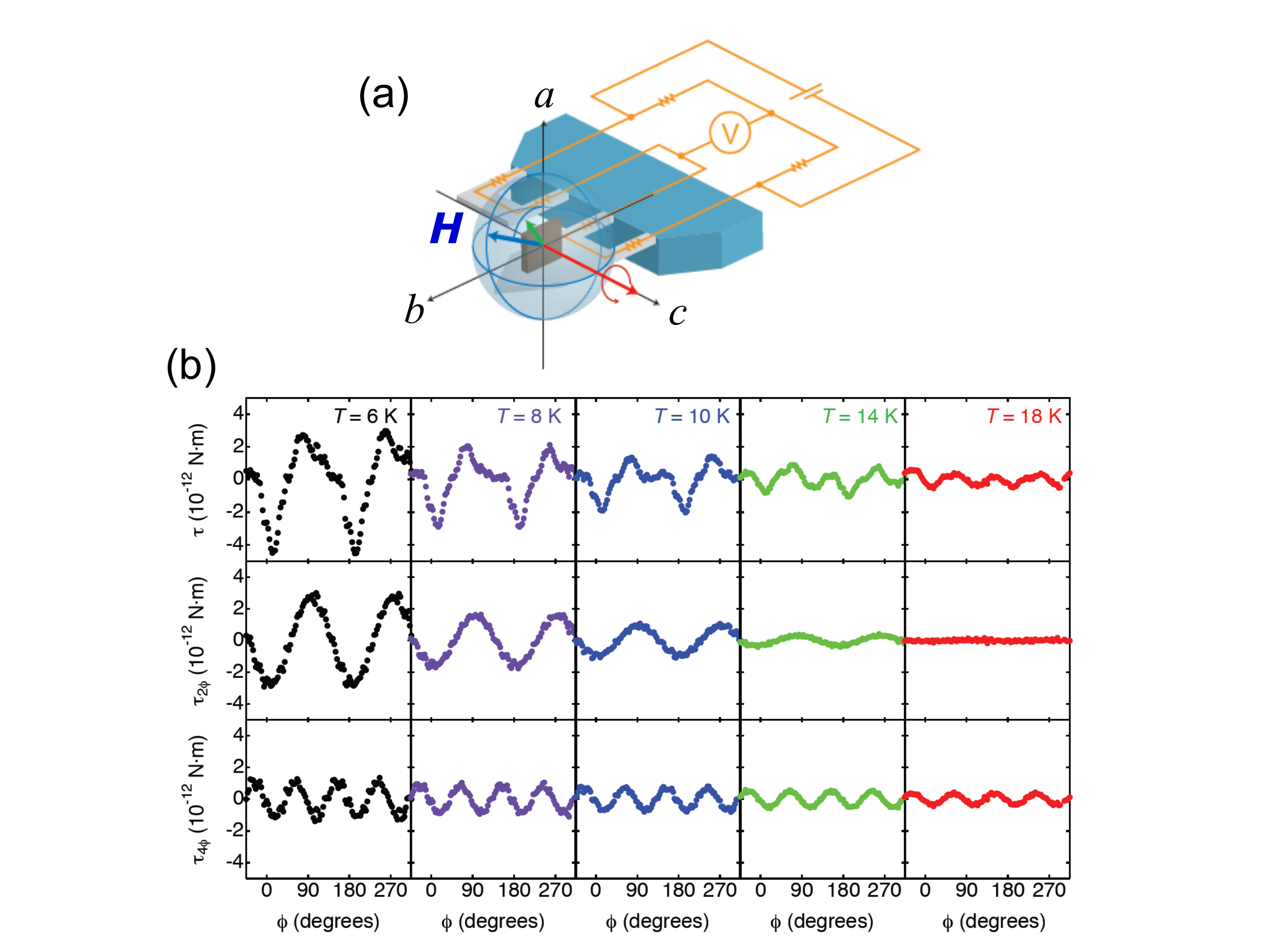}
\caption{(a) Schematic of torque set-up with field rotation in the basal plane and torque in $c$-direction. (b) Top panel: Measured angular dependence of the torque $\tau$ for several temperatures; Middle and bottom panels: decomposition of the torque in $2\varphi$ and $4\varphi$ contributions. After Okazaki et al.\ \cite{okaz11} and Shibauchi and Matsuda \cite{shib12}.}
\label{fig9}
\end{center}
\end{figure}

\begin{figure}
\begin{center}
\includegraphics[width=3in]{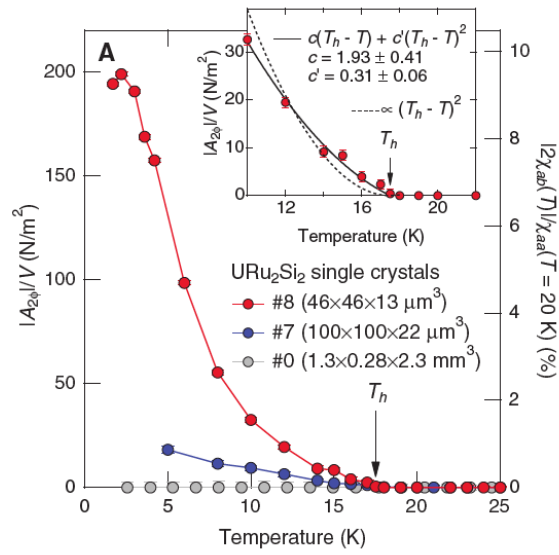}
\caption{Torque amplitude per volume (m$^2$) vs.\ temperature measured on URu$_2$Si$_2$ single crystals of different size \cite{okaz11,shib12}. The amplitude should be proportional to the volume of the crystal. However, the large single crystal shows no torque effect. Only the tiny crystal exhibits the clear upturn below T$_0$. This behaviour has led to the proposal of $[110]$ and $[1\bar{1}0]$ domains in the HO \cite{okaz11}.} 
\label{fig10}
\end{center}
\end{figure}

A {remarkable} series of ``never-tried-before''  experiments on HF materials were carried out by the Kyoto group on URu$_2$Si$_2$ \cite{okaz11,tone12,shib12,tone13}. The first {one} was torque measurements, $\boldsymbol{\tau} = \boldsymbol{M} \times \boldsymbol{H}$, obtained by rotating an applied basal-plane magnetic field $\boldsymbol{H}$ and tracking the {torque $\boldsymbol{\tau}$ Êof the} field-induced magnetisation $\boldsymbol{M} = \boldsymbol{\chi} \cdot \boldsymbol{H}$ in the basal plane as a function of angle, $\varphi$, see Fig.\ \ref{fig9}. {If there is no off-diagonal susceptibility $\chi_{ab}$}, $\boldsymbol{M}\, ||\, \boldsymbol{H}$ and $\boldsymbol{\tau}$ = 0, but {here} there is indeed {a crystal-symmetry breaking}  anisotropy {observed}. When a full rotation 360$^{\circ}$ was accomplished the expected 4-fold symmetry of the {tetragonal} lattice ({two equivalent orthonormal axes of} anisotropy) changed to 2-fold {symmetry} (a single uniaxial anisotropy) in the HO state as detected by the torque oscillations. {Specifically, the torque data reveal rotational symmetry breaking, i.e.\ $\chi_{ab} \ne 0$ in the HO state.} Figure \ref{fig9} also shows  the torque traces as a function of temperature. Note the torque amplitude should be proportional to the volume. In Fig.\ \ref{fig10} the slowly developing 2$\varphi$ torque is displayed as a function of temperature, here the crystal sizes are indicated and the 2$\varphi$ amplitude is normalised per volume in m$^3$.
Figure \ref{fig10} did not have the expected order parameter vs.\ temperature dependence and there was no torque detected for large crystals \cite{okaz11,shib12}.

Fourier analysis of {the} torque curves indicated a 2$\varphi$ oscillation increase as the temperature was lowered below T$_0$. Yet the 4$\varphi$ oscillations persisted and increased into the HO.  Nevertheless  {the conclusion of the torque measurement was} a breaking of 4-fold symmetry, {which was attributed to spin nematicity}. This stimulated {much} theory; {models were developed} along the lines of spin nematic order \cite{fuji11}, multipolar order (specifically, dotriacontapole \cite{iked12},  $E$-type quadrupole \cite{thal11}, antiferro octupole \cite{hanz12}), modulated spin liquid \cite{pepi11}, and circulating spin-charge currents in the basal plane \cite{oppe11} as the cause of fourfold or rotational symmetry breaking. The main conundrum with the experiment lay in the torque magnitude (not $\propto$ volume) only being found in tiniest of crystals. This behaviour then postulated a domain effect that required 50\% of the $\approx$ micron-size domains to be aligned along [110] and the other 50\% along [1$\bar{1}$0], thus cancelling in all but the smallest crystals \cite{okaz11}. However, the question remains: what is causing the domains, a return back to two sources of uniaxial anisotropy? Further, how can the domains be detected on such a fine scale? With fine beam-size synchrotron radiation (50$\mu$m) resonant elastic magnetic and crystalline Bragg-peak diffraction is presently being attempted to determine if the domains are related to the {appearance of} ``puddles" of LMAF regions \cite{xxxx14}.

In a second attempt to validate the two-fold symmetry and domain formation, cyclotron resonance experiments (for the first time on a HF material) were performed on URu$_2$Si$_2$ in the HO state \cite{tone12,tone13}. Cyclotron resonance (CR) measures the effective mass $m_{CR}^*$ of the  conduction electrons moving along extremal orbits of the FS. Here $m_{CR}^* = eH_{CR}/\omega$ and the various electron pockets can be established by determining the resonance lines of $m_{CR}^*$ as a function of frequency, field and angle, thereby enabling a  comparison with QO and DFT.  {Although CR and QO do not yield identical masses, a reasonably} good agreement was first established between CR and QO for the existence of a few large volume pockets, and also those predicted from DFT \cite{tone12}. Figure 16 shows $m _{CR}^*$ for various pockets as the field is rotated from the $c$-direction into the basal plane, and around the basal plane, and finally back to ${c}$ [001]. 

A particular sharp resonant line is the $\alpha$ branch. By field rotation of this branch one can clearly follow its splittings. In the basal plane, [100] to [110],  CR showed a sharp and robust two-fold splitting of the $\alpha$ line moving from [100] to [110] where the largest scattering rate occurred.  {Note that this branch was denoted as $\varepsilon$ in other works \cite{oppe10}.} Moreover, for the QO there was a three-fold splitting that disappeared with a small field rotation out of the basal plane. By taking [110] as a ``hot spot", possible a {FS} gapping direction, a detailed study of the $\alpha$ pocket for all directions through [110] was performed. The $m_{CR}^*$ resonances showed an increase at 45$^{\circ}$ ([110]) and 135$^{\circ}$ ([1$\bar{1}$0]). Here one would not expect the second increase if only [1$\bar{1}$0] domains were present since the cyclotron orbit does not intersect the [110] hot spots. This, thus, is the argumentation for coexistence of distinctly differently oriented domains: (i) the two-fold $\alpha$-band splitting and (ii) the distinct resonance at [1$\bar{1}$0]. So here a dispute exists between QO and CR which must be ironed out by the experts of these two experimental methods.

Presently there is {further} controversy regarding the question: can XRD detect the lattice distortions related to the {proposed} four- to twofold symmetry breaking. The Kyoto team using  {high-resolution X-ray diffraction} of a synchrotron beam at high angles has found a weak indication of a Bragg peak splitting into the HO phase, {expressed as an orthorhombicity of the order of $10^{-5}$} \cite{shib14}. The Sapporo group, {however,} using a more sensitive backscattering synchrotron method has found no signature of lattice distortions {$\Delta a/a$ down to a lower limit of $3 \times 10^{-5}$} \cite{amit14}. As above, there is {sufficient} disagreement between the two different XRD methods to extract {conclusive evidence for a symmetry-breaking} lattice effect.

Another experimental technique brought to bear on the HO symmetry breaking and putative domain formation was $^{29}$Si {nuclear magnetic resonance} (NMR). The JAEA-Tokai group \cite{kamb13} using $^{29}$Si spectra measured the Knight shift and the {NMR} linewidth with basal-plane field rotation {over angle} $\theta$, both above and below T$_0$. NMR should be insensitive to the domains since it probes microscopically the local nuclear sites. By assuming \textit{a priori} the 2-fold symmetry breaking and domains of [110] and [1$\bar{1}$0] orientation, an analysis was drawn with the NMR data. The Knight shift remained constant with $\theta$, and thus was unable to detect 2-fold symmetry or domains. However, the angular variation of the linewidths, Lorentzian fitted, displayed a trough at 45$^{\circ}$ below 10 K. This was explained to arise from the presence of
a domain structure at [110]. Further analyses revealed similar ``sharp minima" at $\theta = 45(1 + 2n)^{\circ}$ where 135$^{\circ}$ corresponding to [1$\bar{1}$0]. The amplitude of these linewidth oscillations increases linearly into the HO state with a $\cos$(2$\theta$) dependence, i.e., a 4-fold symmetry of the troughs.  When this amplitude was compared to that estimated from a mono-domain torque susceptibility measurement \cite{okaz11}, the NMR macroscopic susceptibility {anisotropy} was found to be 15 times less. Kambe et al.'s \cite{kamb13} interpretation was ``the intrinsic twofold  anisotropy ... decreases with increasing sample size" and is not the primary order parameter, {but an accompanying order parameter}.  This then begs the question, after all these excellent experiments, {what are then} the primary order parameter of HO and {its} symmetry breaking. 


\section{New ``exotic" experiments, relation to theory}

We list a series of on-going experiments which need to be fully confirmed and interpreted.
First, there arises the question, does the HO state break time-reversal symmetry (TRS)? The {recent} NMR experiments of Takagi et al.\ \cite{tagi07} {as well as older NMR measurements of Bernal et al.\ \cite{bern06}} and muon spin-rotation measurements of Amato et al. \cite{amat04} say yes. A tiny linewidth broadening is still found on the best crystals; although this broadening is greatly reduced from Bernal's initial NMR results \cite{bern01}. On the other hand,  {recent} very sensitive polar Kerr effect (PKE) measurements of Kapitulnik et al.\ \cite{kapi14} cannot find any magnetic evidence of TRS breaking (TRSB) on the finest of scales in the HO phase. However, when superconductivity appears (T $\lesssim$\,T$_c$) there is a small PKE rotation. Thus, they claim TRSB occurs due to superconductivity but not in the HO,
{which in itself is an extraordinary discovery.}
{The apparent contradicting conclusions of PKE and NMR measurements might stem from the fact that PKE would trace a net ferromagnetic magnetisation, whereas NMR tracks a local magnetisation, which can be in a staggered compensating arrangement.} {Among the theories, TRSB is a clear division line. Multipolar theories such as quadrupolar or hexadecapolar order say no to TRSB, whereas octupolar or dotriacontapolar say yes.}  The recent hastatic {order} theory \cite{chan13} is based upon double TRSB.

{Several recent theories \cite{rauu12,iked12}, including}
this hastatic {order} model \cite{chan13} further predict the appearance of an in-basal-plane magnetic moment. {The appearance of a small in-plane moment would explain the torque measurements of Okazaki et al.\ \cite{okaz11}. However,} three most recent neutron diffraction experiments \cite{pdas13,meto13,ross14} do not observe any in-plane magnetism down to a factor of at least 10 less than the theoretically predicted value. Also the bulk magnetisation measurements of Pfleiderer et al.\ \cite{pfle06} within their sensitivity find no in-plane magnetisation. 

New experimental searches for the HO symmetry, order parameter and elementary excitations include {electronic} Raman scattering \cite{blum14} where a unique A$_{2g}$ mode has been detected and is presently under interpretation. Resonance ultrasonics \cite{ultr14} has detected the lattice softening and HO transition in certain elastic moduli and {is} awaiting a model description.  And finally low-field ESR has been unable to find a resonance mode \cite{esrr14}. Accordingly, the experimental efforts continue unabated with explorations of new methods that offer genetic proving grounds for {strongly correlated electron systems, i.e.} far beyond the HO problem. {Thus far,} most experiments have only a tenuous relationship to a given theory, so as yet the proper theory has not been confirmed by a collection of experiments to the satisfaction of the HO community.

\section{Localizes vs.\ itinerant description of Ising anisotropy}
 
Actinide $5f$ electrons exhibit a dual nature, i.e., they can be atomiclike and localised or hybridised and bandlike. Following this dichotomy  numerous proposed theories are  based on the assumption of localised $5f$ behaviour from the outset. In contrast, other, different theories start from the assumption of itinerant $5f$ electrons. Although somewhat disputed, most of the experimental data appeared thus far to be compatible with delocalised $5f$ character in {\urs}.
Specifically CEF excitations archetypical of localised $5f$ levels could not be detected (see \cite{myds11} for a discussion). 
However, recent QO experiments reported evidence for localised $5f$ character \cite{alta12}. For one of the QO branches 
 a $g$-factor anisotropy ($g_c/g_a$) exceeding 30 was estimated. This implies that the near-Fermi energy quasiparticles would exhibit a giant Ising anisotropy. Such Ising anisotropy would conceivably support the picture of localised $5f$ states in a CEF. Conversely,  for bandlike electrons, in general, a $g$-factor of two with little anisotropy is expected and would definitely seem to disqualify itinerant $5f$ behaviour \cite{alta12}.
 
\begin{figure}
\begin{center}
\includegraphics[width=3in]{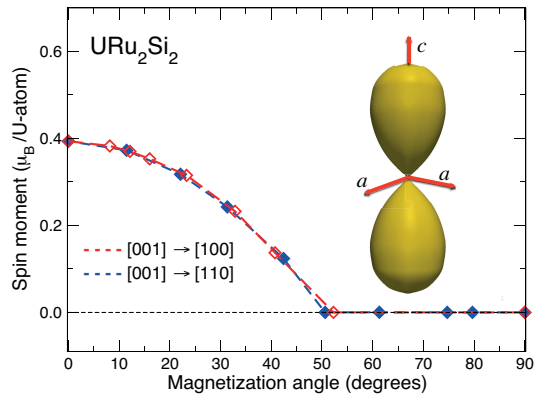}
\caption{Ising anisotropy in {\urs} from DFT calculations. The spin magnetic moment is given as a function of the polar angle $\theta$, for two directions in the unit cell, $[001] \rightarrow [100]$ and $[001] \rightarrow [110]$. The inset shows the magnitude of the computed spin moment as function of its direction in 3D. After Werwi{\'n}ski et al. \cite{werw14}} 
\label{fig11}
\end{center}
\end{figure}

The Ising anisotropy is an unique property of {\urs} whose importance to the HO had previously not been sufficiently realised. This extreme magnetic anisotropy is a central element of the hastatic order theory in which a local $5f^2$ doublet induces the Ising character \cite{chan13}. This scenario of a localised CEF doublet explains the Ising behaviour, but for other CEF schemes, e.g., the two CEF singlets invoking hexadecapolar order \cite{sant94,haul09}, Ising anisotropy is unproven. Surprisingly recent relativistic DFT calculations \cite{werw14} showed that the bandlike $5f$ electrons in {\urs} exhibit a colossal Ising behaviour, a property which is truly exceptional for itinerant electrons. In Fig.\ \ref{fig11} the calculated Ising anisotropy of the total moment on one of the U atoms  is shown; it vanishes steeply for directions away from the $c$-axis. The origin of the unique Ising anisotropy is a combination of the peculiar FS nesting and the strong spin-orbit interaction of uranium. Hence, both the localised $5f^2$ doublet \cite{ohka99,chan13} and DFT band-structure model can explain the Ising anisotropy; for other models of HO such must be demonstrated. 

\section{A confrontation of recent models}

By surveying the remaining viable theories, it is evident that although the hastatic order model can explain the Ising anisotropy, 
the predicted in-plane magnetic moment  could not be observed. 
If the $5f^2$ CEF multiplets would instead consist of two singlets, there would still be the possibility of having hexadecapolar order as proposed by several authors \cite{haul09,kusu11}.  As mentioned above, XRS is possibly the only way to definitely approve or disprove this rank-4 nonmagnetic multipole.
As already mentioned, the recent neutron diffraction analysis \cite{khal14} gave no evidence for the presence of any magnetic multipole.

Several recent ``itinerant models" assume bandlike $5f$ states near the Fermi energy, in conjunction with a FS instability mechanism that causes a FS gapping \cite{elga09,oppe11,dubi11,tdas12,rise12,pepi11,iked12,rauu12}. As mentioned above, the in-plane magnetic moment predicted by Ikeda et al. \cite{iked12} and Rau and Kee \cite{rauu12} could not be confirmed by neutron measurements. Another distinction, which comes into the play, is what the nesting vector is which drives the FS instability. The models of Dubi and Balatsky \cite{dubi11} and Das \cite{tdas12} adopt the incommensurate spin-fluctuations wave vector $\boldsymbol{Q}_1$ as decisive nesting vector, whereas several other models are based on the commensurate wave vector $\boldsymbol{Q}_0$ as the key folding vector \cite{elga09,oppe11,iked12,rauu12}. Note that the model of Riseborough et al.\ \cite{rise12} only assumes that there exists an otherwise unspecified nesting vector. Meanwhile in 2014 it has been demonstrated through various experiments (ARPES, QO) \cite{hass10,meng13} that in the HO phase the body-centred translational symmetry is broken, i.e. there is commensurate $\boldsymbol{Q}_0$ folding of the Brillouin zone in the HO phase. Evidently, in the neighbouring LMAF phase both TRS and the body-centred translation symmetry are broken. As no distinctions between the HO and LMAF FSs have been detected so far, this might suggests that in HO both TRS and body-centred translation are broken. TRSB in the HO -- though not generally agreed -- would be consistent with recent NMR measurements \cite{tagi07}, but not with PKE measurements \cite{kapi14}. However, one would expect an additional symmetry breaking which distinguishes the HO from the LMAF phase. This could be the proposed small orthorhombicity \cite{shib14} or in-plane rotational symmetry breaking \cite{okaz11};  both have been reported but verifications via independent experiments are necessary.

The order parameters proposed in several of the recent itinerant models \cite{oppe11,tdas12,rise12,iked12} have a generic form, $O \propto \langle \sum_{n_{\alpha}n_{\beta} \boldsymbol{k}}
f^{\dagger}_{n_{\alpha}\boldsymbol{k}} \boldsymbol{\Sigma} f_{n_{\beta}\boldsymbol{k}\pm{\boldsymbol{Q}_0} }\rangle$. Here there is a ``pairing amplitude" between $f$ electrons in the two nested bands indexed by $n_{\alpha \boldsymbol{k}}$ and $n_{\beta \boldsymbol{k}^{\prime}}$ and  $\boldsymbol{\Sigma}$ details the interaction operator. The problem of the HO is thus reduced to two $f$ electrons in two doubly-degenerate nested bands. DFT calculations \cite{oppe11} showed furthermore that the relevant bands are formed from the $j=5/2$ manifold and have only doubly-degenerate $j_z=\pm 5/2$ or $j_z = \pm 3/2$ character
when one considers $\boldsymbol{Q}_0$ nesting (but for nesting over $\boldsymbol{Q}_1$ the relevant states have $j_z=\pm 3/2$ and $j_z=\pm 1/2$ character \cite{tdas12}).
In the case of the dotriacontapole proposed by Ikeda et al. \cite{iked12}, both states are assumed to have only $j_z= \pm 5/2$ character and, as they emphasised, only a rank-5 momentum operator $\boldsymbol{\Sigma}$ can couple between $j_z = +5/2$ and $j_z =-5/2$. 
In the spin-orbit density wave model of T.\ Das \cite{tdas12} the operator describes a simultaneous lowering of the orbital index $\ell \rightarrow \ell-1$ and increase of spin index $\sigma \rightarrow \sigma+1$; thus, there is no net change of the total magnetic moment (spin and orbital part), which would explain why HO is so difficult to detect with modern probes of magnetism.

\begin{figure}
\begin{flushleft}
\includegraphics[width=5in]{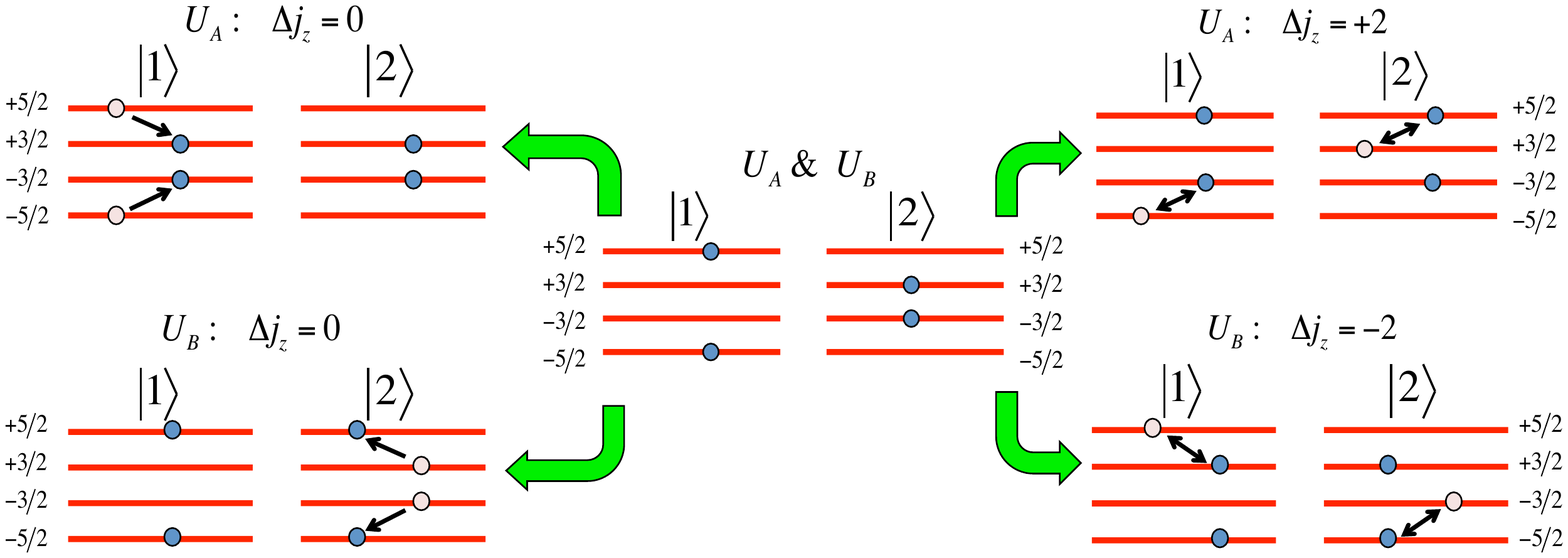}
\vspace*{-3cm}
\caption{Schematics of the two electron rearrangements in the two nested-bands model of {\urs}. The nested band states $|1 \rangle$ and $| 2 \rangle$ have different symmetry $j_z = \pm 5/2$ and $\pm 3/2$, respectively, and because of symmetry protection they cannot hybridise  to open a FS gap. Left side: A two-particle -- two-orbital hopping can occur (cf.\  \cite{rise12}), which is distinct on the two uranium atoms and breaks spontaneously the body-centred lattice symmetry; there is no momentum change on any U atom. Right side: the dynamical symmetry breaking model \cite{oppe11} in which dynamical electron hopping induces an Ising-like excitation with $\Delta j_z =\pm 2$ occurs on one U atom, while simultaneously an antiferromagnetic excitation with $\Delta j_z =\mp 2$ occurs on the other U atom, resulting in a total $\Delta j_z =0$ at all time.} 
\label{fig12}
\end{flushleft}
\end{figure}

The models of Oppeneer et al. \cite{oppe11} and Riseborough et al. \cite{rise12} are both based on a two-electron -- two-orbital rearrangement, schematically shown in Fig.\ \ref{fig12}. Although the model of Riseborough et al.\ \cite{rise12} is originally a spin-only model, it can be reformulated for two states with $j_z$ character. The proposed two-particle rearrangement is such that the total $\Delta j_z =0$, i.e., there is no detectable magnetic moment change. Oppeneer et al.\ \cite{oppe11} assumed dynamical symmetry breaking (i.e., frequency-odd pairing), in which there is an alternating longitudinal moment excitation mode which has an antiferro sense on the two U atoms in the unit cell. One of the U atoms undergoes an oscillating $\Delta j_z=\pm 2$-excitation and the other U atom undergoes an antiparallel $\Delta j_z= \mp 2$ excitation, making the total $\Delta j_z$ vanish at all times (see Fig.\ \ref{fig12}). 
Concomitantly there is a charge-spin current oscillating around each U atom, leading to a nonzero $p_x p_y - p_y p_x$ mode with $A_{2g}$ symmetry.
Validation or rejection of the various recent models is not easy, as this will require dedicated systematic experiments. Such efforts will be required   for finally establishing the nature of the iconic hidden order.

\section{Conclusions}

After nearly three decades of research the origin of the enigmatic HO phase in {\urs} has not yet been unambiguously resolved, at least not to the extent of it being  widely accepted  as to its origin. Nonetheless, in the last few years remarkable progress has been made. Ultra-pure single crystals have been grown, providing more precise data sets than ever before. New experimental techniques have been brought to bear on the problem, leading to a deepened understanding of properties of the HO and superconducting phases and even revealing unexpected aspects., e.g., nodes in the superconducting gap and unusual Raman modes. Also, the last few years saw areas of convergence on features of the HO. For example, Fermi surface cross-sections measured with ARPES \cite{meng13} and with dH-vA and S-dH techniques \cite{ohku99,hass10,alta12}. The developed FS picture is moreover consistent with DFT electronic structure calculations. This implies that, at least the near Fermi-energy states out of which the HO develops are now known. The correspondence between the DFT result and the measured Fermi surface lends credibility to the emerging picture of the HO as a FS reconstruction out of delocalised $5f$ states. Whilst this is gratifying it doesn't yet illuminate the hidden mechanism or driving force leading to the  gap formation. 

Several new possible features of the HO have moved into the forefront. The unexpected discovery of orthorhombicity connected with nematicity in the basal plane \cite{okaz11,shib14} has drawn much attention. Validation of orthorhombicity and nematicity in the HO are key milestones for the near future. Results of other recent measurements such as PKE, elasto-resistivity and electronic Raman scattering are yet to be confronted with current theories of the HO, possibly narrowing further the spectrum of viable models. 
Although this will require further research in the near future, the recently-achieved convergence on current understanding of the HO gives hope that its final unveiling is drawing near.

\section{Acknowledgements}
The Hidden-Order Workshop, held in November 2013 at the Lorentz Center Leiden, The Netherlands, was supported by the Dutch KNAW and ICAM. PMO acknowledges support from the Swedish Research Council (VR).

\newpage

\begin{table}
\caption{
Summary of ongoing contemporary experiments to characterise the heavy fermion precursor, the HO transition and the HO and superconducting states of URu$_2$Si$_2$.}
\begin{tabular}{l l }
\hline \\[-0.2cm]
 Angular resolved photoemission (ARPES) \cite{meng13,boar13,chat13,kawa11,yosh12}   \\
Quantum oscillations (QO) \cite{hass10,alta12,hari13}  \\
Elastic and inelastic neutron scattering  \cite{bour10,ress12,pdas13,meto13,ross14} \\
Nuclear magnetic   and quadrupolar resonance (NMR, NQR)
\cite{tagi07,shir12,kamb13}   \\
Scanning tunneling microscopy (STM) and spectroscopy (STS)
\cite{schm10,ayna10}\\
Ultrafast time-resolved ARPES and reflection spectroscopy \cite{liuu11,dako11}\\
Phononic Raman \cite{meas14} and electronic Raman spectroscopy \cite{blum14} \\
Optical spectroscopy \cite{leva11,guoo12,hall12} \\
Polar Kerr effect \cite{kapi14}  \\
Magnetic torque measurements \cite{okaz11,liii13} \\
Cyclotron resonance \cite{tone12} \\
X-ray diffraction \cite{shib14,amit14} \\
X-ray resonant scattering (XRS) \cite{walk11,walk14} \\
Point contact spectroscopy (PCS) \cite{luuu12,park12,mald12}\\
Resonance ultrasonics \cite{ultr14} \\
Core-level spectroscopy (XAS, EELS) \cite{jeff10} \\
Elasto-resistivity \cite{elas14} \\
 \hline 
\end{tabular}
\end{table}

\begin{table}
\caption{
Summary of analytic theories and models proposed to explain the HO, with an emphasise on the recent contributions. For proposals of specific multipolar magnetic order on the U ions, see Table 3.}
\begin{tabular}{l l }
\hline \\[-0.2cm]
Barzykin \& Gorkov (1995)	& three-spin correlations  \cite{barz95} \\
 Kasuya   (1997) 	&	uranium dimerisation \cite{kasu97} \\
 Ikeda \& Ohashi (1998) 	& d-spin density wave \cite{iked98} \\
 Okuno \& Miyake (1998) &	CEF \& quantum fluctuations \cite{okun98} \\
 Chandra et al.\ (2002) 	&	orbital currents \cite{chan02} \\
 Viroszek et al.\ (2002) 	 &	unconv. spin density wave  \cite{viro02}\\
 Mineev \& Zhitomirsky (2005) 	& staggered spin density wave \cite{mine05} \\
 Varma \& Zhu (2006) 	&	helicity (Pomeranchuk) order \cite{varm06}\\
 Elgazzar et al.\ (2009) 	&	dynamical symmetry breaking \cite{elga09}\\
  Kotetes et al.\ (2010) &	chiral d-density wave \cite{kote10} \\
 Dubi \& Balatsky (2011) 	&	hybridization wave \cite{dubi11}\\
 Pepin et al.\ (2011)	&	modulated spin liquid \cite{pepi11}\\
 Fujimoto (2011)	&	spin nematic order \cite{fuji11} \\
 Riseborough et al.\ (2012)	& unconv. spin-orbital density wave \cite{rise12}\\
 Das  (2012)		&	spin-orbital density wave \cite{tdas12} \\
 Chandra et al.\ (2013)	&	hastatic order \cite{chan13} \\
 Hsu \& Chakravarty (2013) & singlet-triplet d-density wave \cite{hsuu13} \\
\hline 
\end{tabular}
\end{table}

\begin{table}
\caption{
Summary of proposals for a specific multipolar magnetic ordering on the uranium ion to explain the HO, with an emphasise on the recent contributions. Note that different symmetries are possible for high-rank multipoles, therefore some kind of multipoles appear more than once.}
\begin{tabular}{l l }
\hline \\[-0.2cm]
 Nieuwenhuys (1987)     &    dipole  ($2^1$) order \cite{nieu87} \\
Santini \& Amoretti (1994) 	& quadrupolar ($2^2$) order \cite{sant94} \\
 Kiss \& Fazekas (2005) 	&	octupolar ($2^3$) order \cite{kiss05} \\
 Hanzawa \& Watanabe (2005) &	octupolar order  \cite{hanz05}  \\
 Hanzawa (2007)	&	incommensurate octupole \cite{hanz07} \\
 Haule and Kotliar (2009)   & 	hexadecapolar ($2^4$) order \cite{haul09} \\
 Cricchio et al.\ (2009)	 &	dotriacontapolar ($2^5$) order  \cite{cric09} \\
 Harima et al.\ (2010)	 &	antiferro quadrupolar order  \cite{hari10} \\
 Thalmeier \& Takimoto (2011)	& $E(1,1)$-type quadrupole  \cite{thal11}\\
 Kusunose \& Harima (2011)	 &  antiferro hexadecapole\cite{kusu11} \\
 Ikeda et al.\ (2012)    &		$E^-$-type dotriacontapole \cite{iked12} \\
  Rau \& Kee    (2012)       &		$E$-type dotriacontapole \cite{rauu12} \\
 Ressouche et al.\ (2012)  &	dotriacontapolar order  \cite{ress12}\\
\hline 
\end{tabular}
\end{table}

\begin{thebibliography}{43}

\bibitem[1]{pals85}
T.T.M. Palstra, , A.A. Menovsky, J. van den Berg, A.J. Dirkmaat, P.H. Kes, G.J. Nieuwenhuys and J.A. Mydosh, Phys. Rev. Lett. 55 (1985) p.2727.

\bibitem[2]{mapl86}
M.B. Maple, J.W. Chen, Y. Dalichaouch, T. Kohara, C. Rosset, M.S. Torikachvili, M.W. McElfresh and J.D. Thompson, Phys. Rev. Lett. 56 (1986) p.185.

\bibitem[3]{schl86}
W.J. Schlabitz, W.J. Baumann, R. Pollit, U. Rauchschwalbe, H.M. Mayer, U. Ahlheim and C.D. Bredl, Z. Phys. B 62 (1986) p.171.

\bibitem[4]{kasa07}
Y. Kasahara, T. Iwasawa, H. Shishido, T. Shibauchi, K. Behnia, Y. Haga,
T. D. Matsuda, Y. $\bar{\textrm{O}}$nuki, M. Sigrist and Y. Matsuda, Phys. Rev. Lett. {99} (2007) p.116402.

 \bibitem[5]{liii13}
 G. Li, Q. Zhang, D. Rhodes, B. Zeng, P. Goswami, R.E. Baumbach, P.H. Tobash, F. Ronning, J.D. Thompson, E.D. Bauer and L. Balicas, Phys. Rev. B 88 (2013) p. 134517.

\bibitem[6]{myds11}
J.A. Mydosh and P.M. Oppeneer, Rev. Mod. Phys. 83 (2011) p.1301.


\bibitem[7]{meng13}
J.-Q. Meng, P.M. Oppeneer, J.A. Mydosh, P.S. Riseborough, K. Gofryk, J.J. Joyce, E.D. Bauer, Y. Li and T. Durakiewicz, Phys. Rev. Lett. 111 (2013) p.127002.

\bibitem[8]{boar13} 
 F. L. Boariu, C. Bareille, H. Schwab, A. Nuber, P. Lejay, T. Durakiewicz, F. Reinert and A. F. Santander-Syro, Phys. Rev. Lett. 110 (2013) p.156404.

\bibitem[9]{chat13} 
 S. Chatterjee, J. Trinckauf, T. H{\"a}nke, D.E. Shai, J.W. Harter, T.J. Williams, G.M. Luke, K.M. Shen and J. Geck, Phys. Rev. Lett. 110 (2013) p.186401.

\bibitem[10]{kawa11}
I. Kawasaki, S.-i. Fujimori, Y. Takeda, T. Okane, A. Yasui, Y. Saitoh, H. Yamagami, Y. Haga, E. Yamamoto and Y. Onuki, Phys. Rev. B {83} (2011) p.235121.

\bibitem[11]{yosh12}
R. Yoshida, M. Fukui, Y. Haga, E. Yamamoto, Y.  Onuki, M. Okawa, W.
 Malaeb, S. Shin, Y. Muraoka and T. Yokoya, Phys. Rev. B 85 (2012) p.241102(R).
 
\bibitem[12]{hass10}
E. Hassinger, G. Knebel, T.D. Matsuda, D. Aoki, V. Taufour and J. Flouquet,  Phys. Rev. Lett. 105 (2010) p.216409.

\bibitem[13]{alta12}
M.M. Altarawneh, N. Harrison, G. Li, L. Balicas, P.H. Tobash, F. Ronning and E.D. Bauer, Phys. Rev. Lett. 108 (2012) p.066407.

\bibitem[14]{hari13}
N. Harrison, P.J.W. Moll, S.E. Sebastian, L. Balicas, M.M. Altarawneh, J.-X. Zhu, P.H. Tobash, F. Ronning, E.D. Bauer and B. Batlogg, Phys. Rev. B 88 (2013) p.241108(R).

\bibitem[15]{bour10}
F. Bourdarot, E. Hassinger, S. Raymond, D. Aoki, V. Taufour, L.-P. Regnault and J. Flouquet,
J. Phys. Soc. Jpn. 79 (2010) p.064719.

\bibitem[16]{ress12}
E. Ressouche, R. Ballou, F. Bourdarot, D. Aoki, V. Simonet, M.T. Fernandez-Diaz, A. Stunault and J. Flouquet, Phys. Rev. Lett. 109 (2012) p.067202.

\bibitem[17]{pdas13}
P. Das, R. Baumbach, K. Huang, M. Maple, Y. Zhao, J.S. Helton, J. Lynn, E. Bauer and M. Janoschek, New J.Phys. 15 (2013) p.05303.

\bibitem[18]{meto13}
N. Metoki, H. Sakai, E. Yamamoto, N. Tateiwa, T. Matsuda and Y. Haga, J. Phys. Soc. Jpn. 82 (2013) p.055004.

\bibitem[19]{ross14}
K.A. Ross, I. Harriger, Z. Yamani, W.J.L. Buyers, J.D. Garrett, A.A. Menovsky, J.A. Mydosh and C.L. Broholm, Phys. Rev. B 89 (2014) p.155122.

\bibitem[20]{kamb13}
S. Kambe, Y. Tokunaga, H. Sakai, T.D. Matsuda, Y. Haga, Z. Fisk and R.E. Walstedt, Phys. Rev. Lett. 110 (2013) p.246406.

\bibitem[21]{tagi07}
S. Takagi, S. Ishihara, M. Yokoyama and H. Amitsuka,  J. Phys. Soc. Jpn. 81 (2012) p.114710.

\bibitem[22]{shir12}
K.R. Shirer, J.T. Haraldsen, A.P. Dioguardi, J. Crocker, N. apRoberts-Warren, A.C. Shockley, C.-H. Lin, D.M. Nisson, J.C. Cooley, M. Janoschek, K. Huang, N. Kanchanavatee, M.B. Maple, M.J. Graf, A.V. Balatsky and N.J. Curro, Phys. Rev. B 88 (2012) p.094436.

\bibitem[23]{ayna10}
P. Aynajian, E.H. da Silva Neto, C.V. Parker, Y.-K. Huang, A. Pasupathy, J.A. Mydosh and A. Yazdani, Proc. Natl. Acad. Sci. 107 (2010) p.10383.

\bibitem[24]{schm10}
A.R. Schmidt, M.H. Hamidian, P. Wahl, F. Meier, A.V. Balatsky, J.D. Garrett, T.J. Williams, G.M. Luke and J.C. Davis, Nature 464 (2010) p.570.

\bibitem[25]{liuu11}
M.K. Liu and R.D. Averitt, T. Durakiewicz, P.H. Tobash, E.D. Bauer, S.A. Trugman, A.J. Taylor and D.A. Yarotski, Phys. Rev. B 84 (2011) p.161101(R).

\bibitem[26]{dako11}
G.L. Dakovski, Y. Li, S.M. Gilbertson, G. Rodriguez, A.V.
Balatsky, J.-X. Zhu, K. Gofryk, E.D. Bauer, P.H. Tobash, A.J. Taylor, J.L. Sarrao, P.M. Oppeneer, P.S. Riseborough, J.A. Mydosh and T. Durakiewicz, Phys. Rev. B 84 (2011) p.161103(R).

\bibitem[27]{meas14}
M.-A. M{\'e}asson et al., to be published (2014).

\bibitem[28]{blum14}
G. Blumberg et al., to be published (2014).

\bibitem[29]{leva11}
J. Levallois, F. Levy-Bertrand, M.K. Tran, J.A. Mydosh, Y.-K. Huang and D. van der Marel, Phys. Rev. B 84 (2010) p.184420.

\bibitem[30]{guoo12}
W.T. Guo, Z.G. Chen, T.J. Williams, J.D. Garrett, G.M. Luke and N.L. Wang, Phys. Rev. B 85 (2012) p.195105.

\bibitem[31]{hall12}
J.S. Hall, U. Nagel, T. Uleksin, T. Room, T. Williams, G. Luke and T. Timusk, Phys. Rev. B 86 (2012) p.035132. 

\bibitem[32]{kapi14}
A. Kapitulnik et al., to be published (2014)

\bibitem[33]{okaz11}
R. Okazaki, T. Shibauchi, H. J. Shi, Y. Haga, T. D. Matsuda, E. Yamamoto, Y. Onuki, H. Ikeda and Y. Matsuda, Science 331 (2011) p.430.

\bibitem[34]{tone12}
S. Tonegawa, K. Hashimoto, K. Ikada, Y.-H. Lin, H. Shishido, Y. Haga, T.D. Matsuda, E. Yamamoto, Y. Onuki, H. Ikeda, Y. Matsuda and T. Shibauchi, Phys. Rev. Lett. 109 (2012) p.036401.

\bibitem[35]{shib14}
S. Tonegawa, T. Shibauchi, Y. Matsuda, et al., to be published (2014).


\bibitem[36]{amit14}
H. Amitsuka et al. to be published (2014). 

\bibitem[37]{walk11}
 H.C. Walker, R. Caciuffo, D. Aoki, F. Bourdarot, G.H. Lander and J. Flouquet, Phys. Rev. B {83}, (2011) p.193102.

\bibitem[38]{walk14}
H.C. Walker, private communication (2014).

\bibitem[39]{luuu12}
X. Lu, F. Ronning, P.H. Tobash, K. Gofryk, E.D. Bauer and J.D. Thompson, Phys. Rev. B {85} (2012) p.020402(R).

\bibitem[40]{park12}
W.K. Park, P.H. Tobash, F. Ronning, E.D. Bauer, J.L. Sarrao, J.D. Thompson and L.H. Greene, Phys. Rev. Lett. {108} (2012) p.246403.

\bibitem[41]{mald12}
A. Maldonado, I. Guillamon, J. G. Rodrigo, H. Suderow, S. Vieira, D. Aoki and J. Flouquet,
Phys. Rev. B 85 (2012) p.214512. 

\bibitem[42]{ultr14}
A. Shekhter and B. Ramshaw, private communication (2014).

\bibitem[43]{jeff10} 
J.R. Jeffries, K.T. Moore, N. P. Butch and M.B. Maple,
Phys. Rev. B 82 (2010) p.033103.

\bibitem[44]{elas14}
I. Fisher et al., to be published (2014).


\bibitem[45]{barz95}
V. Barzykin and L.P. Gor'kov, Phys. Rev. Lett. {74} (1995) p.4301.

\bibitem[46]{kasu97}
T. Kasuya,  J. Phys. Soc. Jpn. {66} (1997) p.3348.

\bibitem[47]{iked98}
H. Ikeda and Y. Ohashi, Phys. Rev. Lett. {81} (1998) p.3723.

\bibitem[48]{okun98}
Y. Okuno and K. Miyake, J. Phys. Soc. Jpn. {67} (1998) p.2469.

\bibitem[49]{chan02}
P. Chandra, P. Coleman, J. A. Mydosh and V. Tripathi, Nature {417} (2002) p.831.

\bibitem[50]{viro02}
A. Viroszek, K. Maki and B. D{\'o}ra, Int. J. Mod. Phys. B {16} (2002) p.1667.

\bibitem[51]{mine05}
V.P. Mineev and M.E. Zhitomirsky, Phys. Rev. B {72} (2005) p.014432.

\bibitem[52]{varm06}
C.M. Varma and L. Zhu, Phys. Rev. Lett. {96} (2006) p.036405.

\bibitem[53]{elga09} 
S. Elgazzar, J. Rusz, M. Amft, P.M. Oppeneer and J.A. Mydosh, Nat. Mater. {8} (2009) p.337.

\bibitem[54]{kote10}
P. Kotetes,  A. Aperis and G. Varelogiannis, 
ArXiv:cond-mat/1002.2719 (2010).

\bibitem[55]{dubi11} 
Y. Dubi and A.V. Balatsky, Phys. Rev. Lett.  {106} (2011) p.086401. 

\bibitem[56]{pepi11}
C. P{\'e}pin, M.R. Norman, S. Burdin and A. Ferraz, Phys. Rev. Lett. {106} (2011) p.106601.

\bibitem[57]{fuji11}
S. Fujimoto, Phys. Rev. Lett. 106 (2011) p.196407.

\bibitem[58]{rise12} 
P.S. Riseborough, B. Coqblin and S.G. Magalhaes, Phys. Rev. B {85} (2012) p.165116.

\bibitem[59]{tdas12}
T. Das, Sci. Rep. {2} (2012) p.596; see also \cite{tdas14}.

\bibitem[60]{chan13}
P. Chandra, P. Coleman and R. Flint, Nature 493 (2013) p.621.

\bibitem[61]{hsuu13}
C.-H. Hsu and S. Chakravarty, Phys. Rev. B {87} (2013) p. 085114.



\bibitem[62]{nieu87}
G.J. Nieuwenhuys, Phys. Rev. B {35} (1987) p.5260.

\bibitem[63]{sant94}
P. Santini and G. Amoretti, Phys. Rev. Lett. {73} (1994) p.1027.

\bibitem[64]{kiss05}
A. Kiss and P. Fazekas, Phys. Rev. B {71} (2005) p.054415.

\bibitem[65]{hanz05}
K. Hanzawa and N. Watanabe,  J. Phys.: Condens. Matter {17} (2005) p.L419.

\bibitem[66]{hanz07}
K. Hanzawa,  J. Phys.: Condens. Matter {19} (2007) p.072202.

\bibitem[67]{haul09} 
K. Haule and G. Kotliar, Nat. Phys. {5} (2009) p.796.

\bibitem[68]{cric09}
F. Cricchio, F. Bultmark, O. Gr{\aa}n{\"a}s and L. Nordstr{\"o}m, Phys. Rev. Lett. {103} (2009) p.107202.

\bibitem[69]{hari10}
H. Harima, K. Miyake and J. Flouquet, J. Phys. Soc. Jpn. {79}, (2010) p.033705.

\bibitem[70]{thal11}
P. Thalmeier and T. Takimoto, Phys. Rev. B {83} (2011) p.165110.

\bibitem[71]{kusu11}
H. Kusunose and H. Harima, J. Phys. Soc. Jpn. {80} (2011) p.084702.

\bibitem[72]{iked12}
H. Ikeda, M. Suzuki, R. Arita, T. Takimoto, T. Shibauchi and Y. Matsuda, Nat. Phys. {8} (2012) p.528.

\bibitem[73]{rauu12}
J.G. Rau and H.-Y. Kee, Phys. Rev. B {85} (2012) p.245112.



\bibitem[74]{khal14}
D.D. Khalyavin, S.W. Lovesey, A.N. Dobrynin, E. Ressouche, R. Ballou and J. Flouquet, J.Phys.: Condens. Matter 26 (2014) p.046003.

\bibitem[75]{sant09}
A.F. Santander-Syro, M. Klein, F.L. Boariu, A. Nuber, P. Lejay and F. Reinert, Nature Phys. 5 (2009) p.637, and to be published (2014).


\bibitem[76]{ohku99}
H. Ohkuni, Y. Tokiwa, K. Sakurai, R. Settai, T. Haga, E. Yamamoto, Y. Onuki, H. Yamagami, S. Takahashi and T. Yanagisawa, Phil. Mag. B 79 (1999) p.1045.

\bibitem[77]{oppe10}
P.M. Oppeneer, J. Rusz, S. Elgazzar, M.-T. Suzuki, T. Durakiewicz and J.A. Mydosh, Phys. Rev. B 82 (2010) p.205103.


\bibitem[78]{shis09}
H. Shishido, K. Hashimoto, T. Shibauchi, T. Sasaki, H. Oizumi, N. Kobayashi,
T. Takamasu, K. Takehana, Y. Imanaka, T.D. Matsuda, Y. Haga, Y. Onuki and Y. Matsuda,
Phys. Rev. Lett. {102} (2009) p.156403.

\bibitem[79]{broh87}
C. Broholm, J.K. Kjems, W.J.L. Buyers, P. Matthews, T.T.M. Palstra, A.A. Menovsky and J.A. Mydosh, Phys. Rev. Lett. 58 (1987) p.1467.

\bibitem[80]{broh91}
C. Broholm, H. Lin. P.T. Matthews, W.J.L. Buyers, M.F. Collins, A.A. Menovsky, J.A. Mydosh and J.K. Kjems, Phys. Rev. B 43 (1991) p.12809.


\bibitem[81]{wieb07}
C.R. Wiebe, J.A. Janik, G.J. MacDougall, G.M. Luke, J.D. Garrett, H.D. Zhou, Y.-J. Jo, L. Balicas, Y. Qiu, J.R.D. Copley, Z. Yamani and W.J.L. Buyers, Nat. Phys. {3} (2007) p.96.


\bibitem[82]{oppe11}
P.M. Oppeneer, S. Elgazzar, J. Rusz, Q. Feng, T. Durakiewicz and J.A. Mydosh, Phys. Rev. B 84 (2011) p.241102R.

\bibitem[83]{tdas14}
T. Das, Phys. Rev. B 89 (2014) p.045135.

\bibitem[84]{kuwa06}
K. Kuwahara, M. Kohgi, K. Iwasa, M. Nishi, K. Nakajima, M. Yokoyama and H. Amitsuka, Physica B 378-380 (2006) p.581


\bibitem[85]{ayna14}
P. Aynajian, E.H. da Silva Neto, B. Zhou, S. Misra, R.E. Baumbach, Z. Fisk, J.A. Mydosh, J.D. Thompson, E.D. Bauer and A. Yazdani, to be published in J. Phys. Soc. Jpn. (2014).

\bibitem[86]{bonn88}
D.A. Bonn, J.D. Garret and T. Timusk, Phys. Rev. Lett. 61 (1988) p.1305.


\bibitem[87]{hara11}
J.T. Haraldsen, Y. Dubi, N.J. Curro and A.V. Balatsky, Phys. Rev.
B 84 (2011) p.214410.

\bibitem[88]{shib12}
T. Shibauchi and Y. Matsuda, Physica C 481 (2012) p.229.

\bibitem[89]{tone13}
S. Tonegawa, K. Hashimoto, K. Ikada, Y. Tsuruhara, Y.-H. Lin, H. Shishido, Y. Haga, T.D. Matsuda, E. Yamamoto, Y. Onuki, H. Ikeda, Y. Matsuda and T. Shibauchi, Phys. Rev. B 88 (2013) p.245131.

\bibitem[90]{hanz12}
K. Hanzawa, J. Phys. Soc. Jpn. {\bf 81}, 114713 (2012).

\bibitem[91]{xxxx14}
P. Aynajian and H.C. Walker, private communication (2014).

\bibitem[92]{bern06}
O.O. Bernal, M.E. Moroz, K. Ishida, H. Murakawa, A.P. Reyes, P.L. Kuhns, D.E. MacLaughlin, J.A. Mydosh and T.J. Gortenmulder, Physica B 378-380 (2006) p.574.

\bibitem[93]{amat04}
A. Amato,  M.J. Graf, A. de Visser, H. Amitsuka, D. Andreica and A. Schenck, 
J. Phys.: Condens. Matter {16} (2004) p.S4403.

\bibitem[94]{bern01}
O.O. Bernal, C. Rodrigues, A. Martinez, H.G. Lukefahr, D.E. MacLaughin, A.A. Menovsky and J.A. Mydosh, Phys. Rev. Lett.  87 (2001) p.196402.

\bibitem[95]{pfle06}
C. Pfleiderer, J.A. Mydosh and M. Vojta, Phys. Rev. B 74 (2006) p.104412.

\bibitem[96]{esrr14}
J. Sichelschmidt, private communication (2013).

\bibitem[97]{werw14}
M. Werwi{\'n}ski, J. Rusz, J.A. Mydosh and P.M. Oppeneer, to be published (2014).

\bibitem[98]{ohka99}
F. J. Ohkawa and H. Shimizu, J. Phys. Condens. Matter {11} (1999) p.L519.




\end{thebibliography}
\end{document}